\documentclass{cja}

\usepackage{enumerate}
\usepackage{natbib}

\usepackage{bm}
\usepackage{color}
\usepackage{booktabs}
\usepackage{tabularx}
\usepackage{multirow}
\usepackage[figuresright]{rotating}
\usepackage{amsmath}
\usepackage {times}

\usepackage{hyperref}
\usepackage{cleveref}

\begin{document}



\title{Review of X-ray pulsar spacecraft autonomous navigation}




\author[a]{Yidi WANG\corref{cor}}
\ead{wangyidi_nav@163.com}
\author[a]{Wei ZHENG}
\author[b,c]{Shuangnan ZHANG}
\author[b]{Minyu GE}
\author[d]{Liansheng LI}
\author[e,f]{Kun JIANG}
\author[g]{Xiaoqian CHEN}
\author[h]{Xiang ZHANG}
\author[b]{Shijie ZHENG}
\author[b]{Fangjun LU}

\shortauthors{Y.D. WANG and et al.}

\cortext[cor]{Corresponding author}

\affiliation[a]{
  organization = {College of Aerospace Science and Engineering, National University of Defense Technology},
  city         = {Changsha 410073},
  country      = {China},
}

\affiliation[b]{
  organization = {Key Laboratory of Particle Astrophysics, Institute of High Energy Physics, Chinese Academy of Sciences},
  city         = {Beijing 100049},
  country      = {China},
}

\affiliation[c]{
  organization = {University of Chinese Academy of Sciences, Chinese Academy of Sciences},
  city         = {Beijing 100049},
  country      = {China},
}

\affiliation[d]{
  organization = {Beijing Institute of Control Engineering},
  city         = {Beijing 100080},
  country      = {China},
}

\affiliation[e]{
  organization = {Beijing Institute of Tracking and Communication Technology},
  city         = {Beijing 100091},
  country      = {China},
}

\affiliation[f]{
  organization = {Aerospace Information Research Institute, Chinese Academy of Science},
  city         = {Beijing 100091},
  country      = {China},
}
\affiliation[g]{
  organization = {Academy of Military Science},
  city         = {Beijing 100091},
  country      = {China},
}
\affiliation[h]{
  organization = {National Innovation Institute of Defense Technology, Academy of Military Science},
  city         = {Beijing 100091},
  country      = {China},
}

\begin{abstract}
  This article provides a review on X-ray pulsar-based navigation (XNAV). The review starts with the basic concept of XNAV, and briefly introduces the past, present and future projects concerning XNAV. This paper focuses on the advances of the key techniques supporting XNAV, including the navigation pulsar database, the X-ray detection system, and the pulse time of arrival estimation. Moreover, the methods to improve the estimation performance of XNAV are reviewed. Finally, some remarks on the future development of XNAV are provided.
\end{abstract}



\maketitle


\section{Introduction}\label{sect:intro}

Advances in the aerospace engineering significantly boost the deep space explorations. The Deep Space Networks (DSNs) can track deep space spacecraft by measuring the ranges and range rates of spacecraft with respect to the ground stations. In this case, the radial positions of spacecraft can be accurately determined, but the components of positions perpendicular to the spacecraft-Earth line have large errors \citep{Becker2013}. Typically, the error of positions perpendicular to the spacecraft-Earth line is around 4 km per Astronomical Unit (AU) of distance between the Earth and spacecraft \citep{James2009}. Finally, the uncertainty will grow to a level of the order of ±200 km at the orbit of Pluto and $\pm500$ km at the orbit of Voyager 1 \citep{Becker2013}. Therefore, for a deep space spacecraft, it is better to determine the position and velocity autonomously using onboard measurements. The optical-imaging-based autonomous navigation system is well-developed, and has been applied to various deep space missions \citep{Williams2002}. However, this method is effective only when a spacecraft is planning to land on a celestial body \citep{Mourikis2009, Amzajerdian2011, GASKELL2008}. Currently, there lacks an autonomous navigation system for a cruising spacecraft.

X-ray pulsar navigation (XNAV) system is a promising autonomous navigation system for spacecraft traveling through the solar system \citep{Wood1993SPIE}. XNAV employs the X-ray radiation from pulsars to estimate the position and velocity of a spacecraft. A pulsar is a rapidly spinning neutron star, a product of a massive star coming to the end of its lifetime \citep{Lorimer2004}, and can emit the electromagnetic radiation at regular intervals. The first pulsar was discovered in 1967 \citep{Hewish1969}. A spacecraft receives electromagnetic signals from pulsars just like a ship receiving signals from a lighthouse on the coast. It is the very reason why pulsars are also called the "lighthouses in the universe" \citep{Lattimer2004}. All the pulsars locate far from the solar system, and their positions can be measured in advance. Then, all the pulsars compose an interstellar navigation system. If a spacecraft receives pulsed signals from different pulsars, its position and velocity can be estimated well, just like a vehicle that can autonomously determine its position and velocity by receiving signals from Global Navigation Satellite System (GNSS). It is not a new idea. Voyager 1 and Voyager 2 spacecraft were equipped with the Golden Voyager Record, which marked the position of the Earth with the aid of 14 pulsars \citep{kohlhase1977voyager}. The Pioneer plaques, fixed to the Pioneer 10 and Pioneer 11, also marked the position of the Earth relative to pulsars \citep{Sagan1972}.

This paper will review the development of XNAV. The remainder of this paper is organized as follows. Section 2 introduces the basic concept of XNAV. The missions concerning XNAV are reviewed in Section 3. Section 4 provides the principle of XNAV. Sections 5-7 review the advances of navigation pulsar database, X-ray detection system and pulse Time of Arrival (TOA) estimation, respectively. The methods to improve the estimation peformance of XNAV are briefly reviewed in Section 8. Section 9 concludes the paper and presents the prospects in future work.

\section{Basic concepts about XNAV}\label{sect:prin_XNAV}
\subsection{Basic concepts of a pulsar}
A star in the mass range of about 8 to 30 times the solar mass ends up with a neutron star. The mass of a neutron star is typically 1.4-3.0 solar mass, and the radius of a neutron star is only about 10 km \citep{Becker2013}. Pulsars are rapidly spinning and strongly magnetized neutron stars.

\subsubsection{Category}\label{sect:cate}
According to the energy source of electromagnetic radiation, pulsars can be classified into three categories \citep{Becker2013}:

\begin{enumerate}[(1)]
  \item Accretion-powered pulsars\\
  Accretion-powered pulsars are within binary systems, which are usually composed of a companion star and a neutron star. The neutron star accretes the matter from the companion star. However, accretion-powered pulsars have a complex spin frequency evolution, and thus are disqualified as reference sources for navigation \citep{Ghosh2007}.
  \item Magnetars\\
  Magnetars are isolated neutron stars with exceptionally high magnetic dipole fields of up to $10^{15}$ Gauss. The spin frequency evolution of a magnetar is virtually unstable, which hinders magnetars from being reference sources for navigation.
  \item Rotation-Powered Pulsars (RPP)\\
  RPP radiate by consuming the rotation energy of themselves. Most of RPPs can radiate in a broadband (from radio to optical, X-rays and gamma-rays), and are typically recommended for navigation.
\end{enumerate}

According to the spin period, the reciprocal of spin frequency, RPPs can be divided into two types:
\begin{enumerate}[(1)]
  \item Milli-Second Pulsars (MSP)\\
  MSPs have spin period less than 20 milliseconds. The period of the fastest pulsar is about 1.4 milliseconds.
  \item Normal pulsars\\
  Normal pulsars have spin periods in the range of tens of milliseconds to several seconds.
\end{enumerate}

\subsubsection{Naming convention}
The name of a pulsar usually starts with the abbreviation PSR (Pulsating Source of Radio), although pulsars can radiate at other wavelengths of the electromagnetic spectrum meanwhile. PSR is followed by an abbreviation for the epoch, then the pulsar’s right ascension and declination. Pulsars discovered before 1993 use the Besselian epoch. Pulsars discovered after 1993 use the Julian epoch. For example, a pulsar named PSR B1937+21 indicates the pulsar has a right descent of $19\mathrm{h}37\mathrm{m}$ and a declination of 21$^\circ$ relative to the Besselian epoch.

\subsubsection{Stability of spin frequency of pulsar}
Pulsars can be employed to determine the positions of spacecraft, not only because the positions of pulsars can be well determined in advance but also because the spin frequencies of pulsars are quite stable \citep{Rawley1987}. The stability of frequency can be scaled by the Square Root Allan Variance (SRAV). When there are $N$ frequency measurements of a source, $f_{1}, f_{2}, \cdots, f_{N}$, the SRAV is defined as \citep{Hartnett2011}
\begin{equation}
  \sigma_{y}^{2}(\tau)=\frac{1}{2(N-1) \nu_{0}^{2}} \sum_{k=1}^{N-1}\left(f_{k+1}-f_{k}\right)^{2}
\end{equation}
where $\nu_{0}$ is the nominal average frequency of the source.

Ref.\citenum{Hartnett2011} compares the SRAV of MSPs with the current high-precision timing instruments, including optical clocks, caesium clocks and microwave clocks. It is found that the long-term stability of MSPs is comparable to the current timing instruments.

\subsubsection{Pulse profiles of pulsars}
\begin{figure}[h]
  \centering
  \includegraphics[width=\columnwidth]{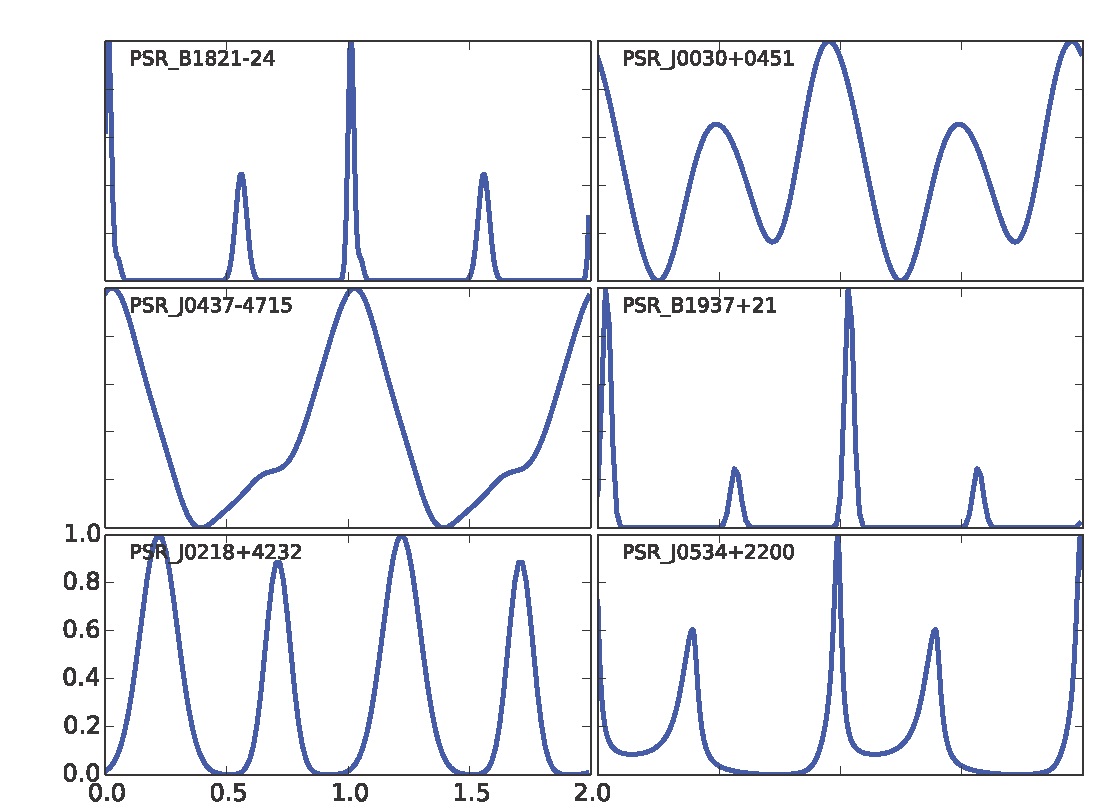}
  \caption{X-ray pulse profiles of pulsars for SEXTANT program\citep{Jason2015}}\label{fig:pulsar_profile}
\end{figure}

The shape of pulsed signals of a pulsar is called profile in the pulsar astronomy community. Each pulsar has a unique profile, just like the fingerprint of a person. Fig.\ref{fig:pulsar_profile}\citep{Jason2015} shows the profiles of pulsars that are selected by the Station Experiment for X-ray Timing and Navigation Technology (SEXTANT) program, which will be introduced in Section \ref{sect:sextant}.

\subsection{Coordinate and time systems for XNAV}
In the aerospace engineering, the orbit motion of spacecraft is described in a coordinate system. XNAV employs the coordinate systems identical to other navigation systems, including (A) the J2000.0 Earth-centered inertial coordinate system for Earth-orbiting spacecraft and (B) the J2000.0 heliocentric-ecliptic coordinate system for deep space spacecraft\citep{Hintz2015}.

Given that pulsars locate far away from the solar system, there are numerous huge celestial bodies between pulsars and spacecraft. Given that the space is warped by the gravity of the celestial bodies, the time system for XNAV should take the general relativity effect into consideration. As a consequence, the time system for XNAV is the Barycentric Dynamic Time (TDB) or Barycentric Coordinate Time (TCB) \citep{IAU2014}.

\section{Past, present and future projects concerning XNAV}
Since the first proposal of pulsar navigation and timing\citep{Downs1974,Chester1981}, there have been XNAV related projects in the US, China, and Europe. This section will briefly review these projects in chronological order.

\subsection{Past projects}
\subsubsection{USA experiment}\label{sect:usa}
\begin{figure}[h]
  \centering
  \includegraphics[width=\columnwidth]{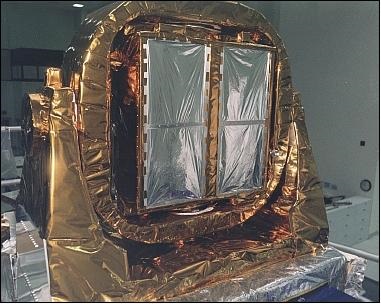}
  \caption{Illustration of USA instrument (image credit: Naval Research Laboratory of the United States)}\label{fig2}
\end{figure}

The Unconventional Stellar Aspect (USA) experiment is one of the nine experiments on board the Air Force Space Test Program's Advanced Research and Global Observation Satellite (ARGOS) launched in 1999, and operated from May 1, 1999, to November 16, 2000. This experiment consisted of a pair of gas scintillation proportional counters to detect X-ray radiations \citep{Ray2001} (see Fig.\ref{fig2}) and a Global Positioning System (GPS) receiver for recording the arrivals of X-ray photons.

The USA experiment diagnosed and corrected a problem with the attitude control of the ARGOS by observing an X-ray source \citep{Kowalski2001, Wood2017}, and tried to fulfill the timekeeping using the observation on Crab pulsar \citep{Wood1993SPIE}.

\subsubsection{XNAV program}
The Defense Advanced Research Projects Agency (DARPA) of the United States started a program, X-ray Source Based Navigation for Autonomous Position Determination, which is also referred to as the XNAV program, in 2004. The XNAV program aimed to develop a revolutionary attitude and navigation capability exploiting periodic celestial sources such as pulsars in the X-ray band \citep{beckett2006}.

This program was conceived to last from 2005 to 2010, consisting of three phases. At the end of the program, a flight experiment XNAV was planned to perform on the International Space Station (ISS). However, the whole program was terminated at the end of Phase I (September 2006). Nevertheless, this program performed an in-depth investigation on the feasibility of XNAV, and realized the following achievements \citep{beckett2006}:
\begin{enumerate}[(1)]
  \item Identifying more than 10 candidate pulsars, whose spin stabilities and X-ray fluxes are capable for precise navigation purpose.
  \item Complementation of X-ray detector developments.
  \item Demonstrating navigation algorithms for spacecraft position/velocity determination.
  \item Designing the demonstration system architecture planned for ISS flight.
\end{enumerate}

\subsubsection{XTIM program}
Between 2009 and 2010, DARPA \citep{SHEIKH2011} performed the X-ray Timing (XTIM) seedling program to study the use of pulsars for time transfer and investigate future demo scenarios. The XTIM was envisioned to create a Pulsar Timescale (PT) by combining long-term pulsar observations with an ultra-stable local clock for short-term stability. The designed XTIM system was expected to have the following potential characteristics \citep{SHEIKH2011}:
\begin{enumerate}[(1)]
  \item Stability: XTIM was based on observations of highly stable pulsars.
  \item Autonomy: XTIM would provide users with an independent and precise measurement of time.
  \item Universality: The pulsars used by XTIM are celestial sources, making them available to any user, anywhere in the solar system.
\end{enumerate}

\subsection{Present missions}
\subsubsection{NICER/SEXTANT program}
\label{sect:sextant}
The Goddard Space Flight Center (GSFC) of the National Aeronautics and Space Administration (NASA) of the US initialized a science mission, Neutron-star Interior Composition Explorer (NICER), in 2011. The fundamental science of NICER is understanding the ultra-dense matter through observations of neutron stars in the soft X-ray band \citep{Gendreau2012S}. Besides the science goal, the NICER mission has a technology demonstration enhancement called Station Experiment for X-ray Timing and Navigation Technology (SEXTANT) \citep{Jason2015}. The object of SEXTANT is to perform a real-time, on-board XNAV-only orbit determination with an accuracy of 10 km in the worst-direction, using up to two weeks of dedicated navigation-focused MSP observations.

\begin{figure}[h]
  \centering
  \includegraphics[width=\columnwidth]{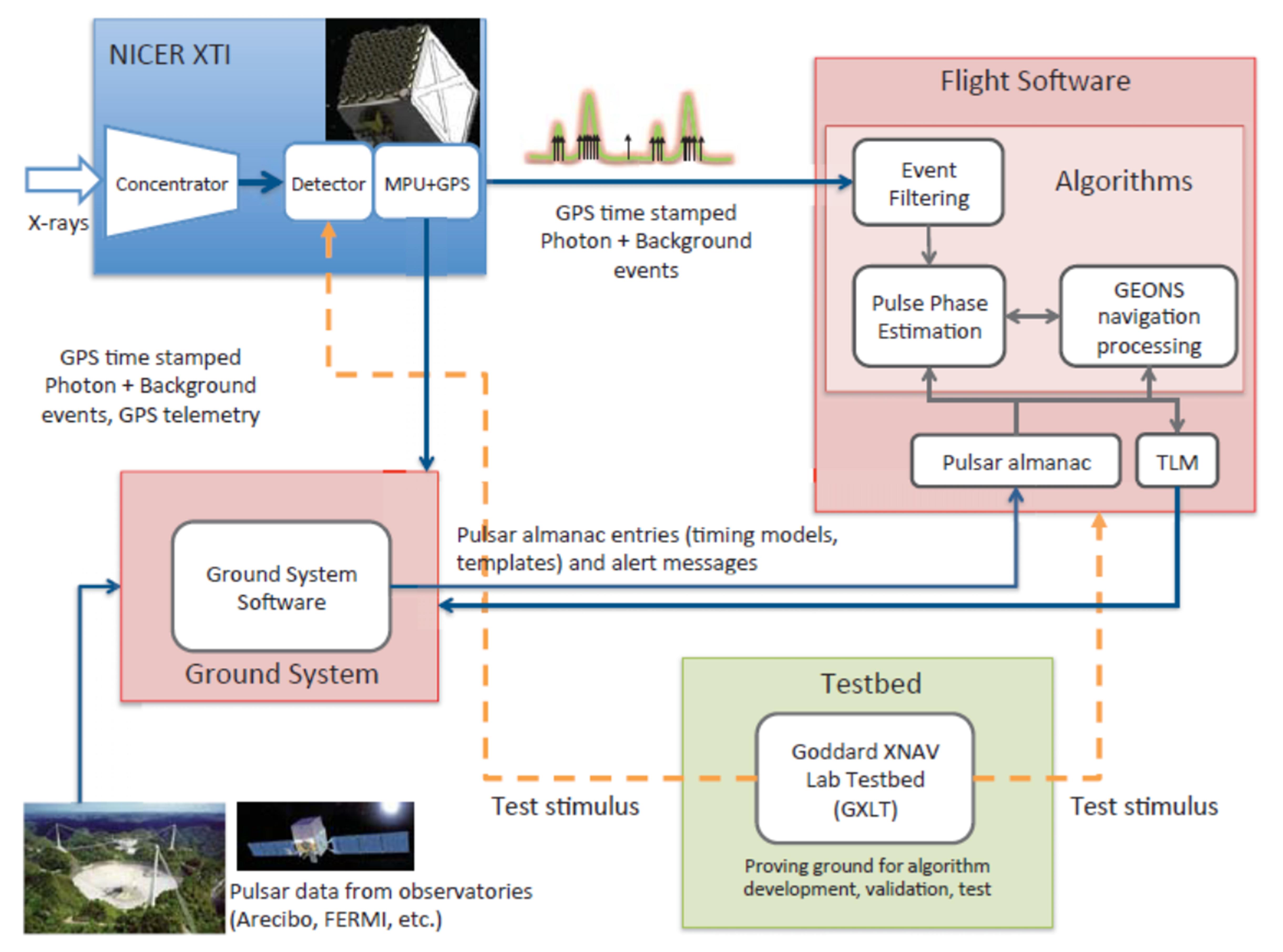}
  \caption{SEXTANT system architecture \citep{Jason2015}}\label{fig4}
\end{figure}

The SEXTANT is the first real-time flight experiment on XNAV in the world. As shown in Fig.\ref{fig4} \citep{Jason2015}, the system architecture of SEXTANT includes four main components: (A) the NICER X-ray Timing Instrument (XTI), (B) the flight software and algorithms, (C) the ground system, and (D) the ground testbed\citep{Mitchell2014}. In 2018, SEXTANT has successfully demonstrated the XNAV aboard the ISS and achieved a positioning error below 10 km \citep{Witze2018}.

\subsubsection{PULCHRON project}
In December 2018, the European Space Agency (ESA) announced a project called PULCHRON, an abbreviation built with the words “Pulsar” and “Chronos”, the ancient Greek term for “Time” \citep{Piriz2019}. The primary object of PULCHRON is to demonstrate the effectiveness of a pulsar time scale for the generation and monitoring of system timing in Positioning, Navigation and Timing (PNT) in general, and of Galileo System Time (GST) in particular.

The PULCHRON project is composed of (A) Pulsar Measurement Collection System (PMCS), (B) Pulsar Frequency Standard Unit (PFSU) and (C) Pulsar-Augmented Clock Ensemble (PACE) \citep{Esteban2019}, where PMCS monthly collects the pulsar measurements from the European Pulsar Timing Array (EPTA), which includes Effelsberg, Lovell, Westerbork, Nançay and Sardinia \citep{Ferdman_2010}; PFSU aims to realize the physical pulsar time scale; and PACE builds a posteriori "paper" time scale mixing pulsar data and GNSS station and satellite clocks.

\subsubsection{XPNAV-1 satellite}
China launched an experiment satellite called X-ray Pulsar navigation-1 (XPNAV-1) in November 2016. The objective of XNAV-1 is to test the technology of pulsar observation in the soft X-ray band with the X-ray instruments developed by China Academy of Space Technology (CAST) \citep{Zhang2017}.

Up to now, the XPNAV-1 has accomplished observations on 4 isolated RPPs and 4 X-ray binaries \citep{Zhang2017}, performed a rough orbit determination with an averaged accuracy about 38.4 km \citep{Huang2019}, and completed a timely observation on the glitch of Crab pulsar \citep{Huang2017}.

\subsubsection{Navigation study of POLAR}
POLAR was a hard X-ray polarimeter aboard the Chinese space laboratory TianGong-2 (TG-2), which was launched in 2016 and deliberately de-orbited in July 2019. It was equipped with an array of 1600 plastic scintillators for precisely and efficiently measuring the linear polarization of hard X-rays. During the mission, POLAR detected 55 $\gamma$-ray bursts.

Besides observing the $\gamma$-ray bursts, POLAR performed a navigation study. The orbit of TG-2 was determined to have an accuracy of about 10 km with 31-day-long hard X-ray data on Crab pulsar. The feasibility of XNAV was validated in principle \citep{Zheng2017SSPMA}.

\subsubsection{Navigation study of Insight-HXMT}
The Insight-Hard X-ray Modulation Telescope (Insight-HXMT), China’s first X-ray astronomy satellite, was launched in June 2017. The main scientific objectives of Insight-HXMT include: (A) scanning the Galactic Plane to find new transient sources and to monitor the known variable sources, (B) observing X-ray binaries to study the dynamics and emission mechanism in strong gravitational or magnetic fields, and (C) monitoring and studying Gamma-Ray Bursts (GRBs) and Gravitational Wave Electromagnetic counterparts (GWEM) \citep{ZhangSN2020}.

Besides accomplishing the scientific objects, Insight-HXMT completed a flight study on XNAV by employing the X-ray data of the Crab pulsar collected between August 30 and September 3, 2017. As a result, it was claimed that Insight-HXMT can locate itself with an error below 10 km (3 $\sigma$)\citep{Zheng_2019, SUN2022}.

\subsection{Future missions}
\subsubsection{CubeX}
The SEXTANT team is working with scientists from Smithsonian Astrophysical Observatory and Harvard
University to test XNAV on a CubeSat mission called CubeX. CubeX plans to employ an X-ray telescope to identify and spatially map lunar crust and mantle materials excavated by impact craters, and also serves as a pathfinder for autonomous precision deep space navigation using X-ray pulsars \citep{Romaine2018}.

\subsubsection{XNavSat}
The Indian Institute of Space Science and Technology is developing a small satellite mission called XNavSat. This mission aims to calculate the true anomaly of the satellite in real time, assuming that the remaining orbital elements of the satellite are known \citep{Anant2021}.

\section{Principle of XNAV}\label{sect:p_XNAV}
\subsection{XNAV for a single spacecraft}\label{sect:p_XNAV_1}
The basic idea of XNAV for a single spacecraft is similar to that of GNSS. When a spacecraft receives signals from a pulsar, the TOAs of the signals, $t_{\mathrm{SC}}$, can be measured.

Given that the spin frequency of pulsar is stable, the evolution of the pulse phase of pulsar at the Solar System Barycenter (SSB) can be described by
\begin{equation}
  \label{eq.timing_model}
  \begin{aligned}
      \phi\left(t_{\mathrm{SSB}}\right)=\phi_{T_{0}}+\sum^{3}_{i=1}\frac{F_{i-1}}{i!}\left(t_{\mathrm{SSB}}-T_{0}\right)^{i}
  \end{aligned}
\end{equation}
where $F_{0}$, $F_{1}$ and $F_{2}$ are the spin frequency of the pulsar at $T_{0}$, its time derivative and its second order time derivative respectively. $\phi_{T_{0}}$ is the phase at $T_{0}$. In practice, $F_{0}$, $F_{1}$ and $F_{2}$ can be found in the public pulsar ephemerides \citep{Alam2021, Lyne1993}.

Eq.(\ref{eq.timing_model}) is essentially a Taylor expansion of the pulse phase around the pulse phase at $T_{0}$. If the high orders of the time derivatives of the spin frequency of a pulsar are taken into consideration, the pulse phase evolution should be described more accurately than Eq.(\ref{eq.timing_model}). However, in practice, the spin frequency and its time derivative families are fitted by the real pulsar data. Up to now, the observation on pulsar has lasted for 55 years \citep{Nanda2017}, and the current pulsar ephemerides only contain the second-order time derivative of spin frequency at most.

In Eq.(\ref{eq.timing_model}), $t_{\mathrm{SSB}}$ is expressed as \cite{Edwards_2006}
\begin{equation}
  \label{BC}
  \begin{aligned}
      t_{\mathrm{SSB}}&=t_{\mathrm{SC}}+\frac{1}{c}\boldsymbol{n}^{\mathrm{T}}\boldsymbol{r}_{\mathrm{SSB}}(t)\\
      &+2 \sum_{k} \frac{G M_{k}}{c^{3}} \ln \left(\boldsymbol{n}^{\mathrm{T}}\boldsymbol{r}_{k}(t)+\left\|\boldsymbol{r}_{k}(t)\right\|\right)
  \end{aligned}
\end{equation}
where $\boldsymbol{r}_{\mathrm{SSB}}(t)$ is the position of the spacecraft relative to the SSB, $\boldsymbol{n}$ denotes the direction vector of the pulsar, $G$ is the gravitational coefficient, $ M_{k}$ is the mass of the $k$th celestial body, $\boldsymbol{r}_{k}(t)$ is its position relative to the spacecraft, $c$ is the speed of light, and $\left\|\bullet\right\|$ denotes the $\ell_{2}$-norm of the vector within.

\begin{figure}[!htp]
  \centering
  \includegraphics[width=\columnwidth]{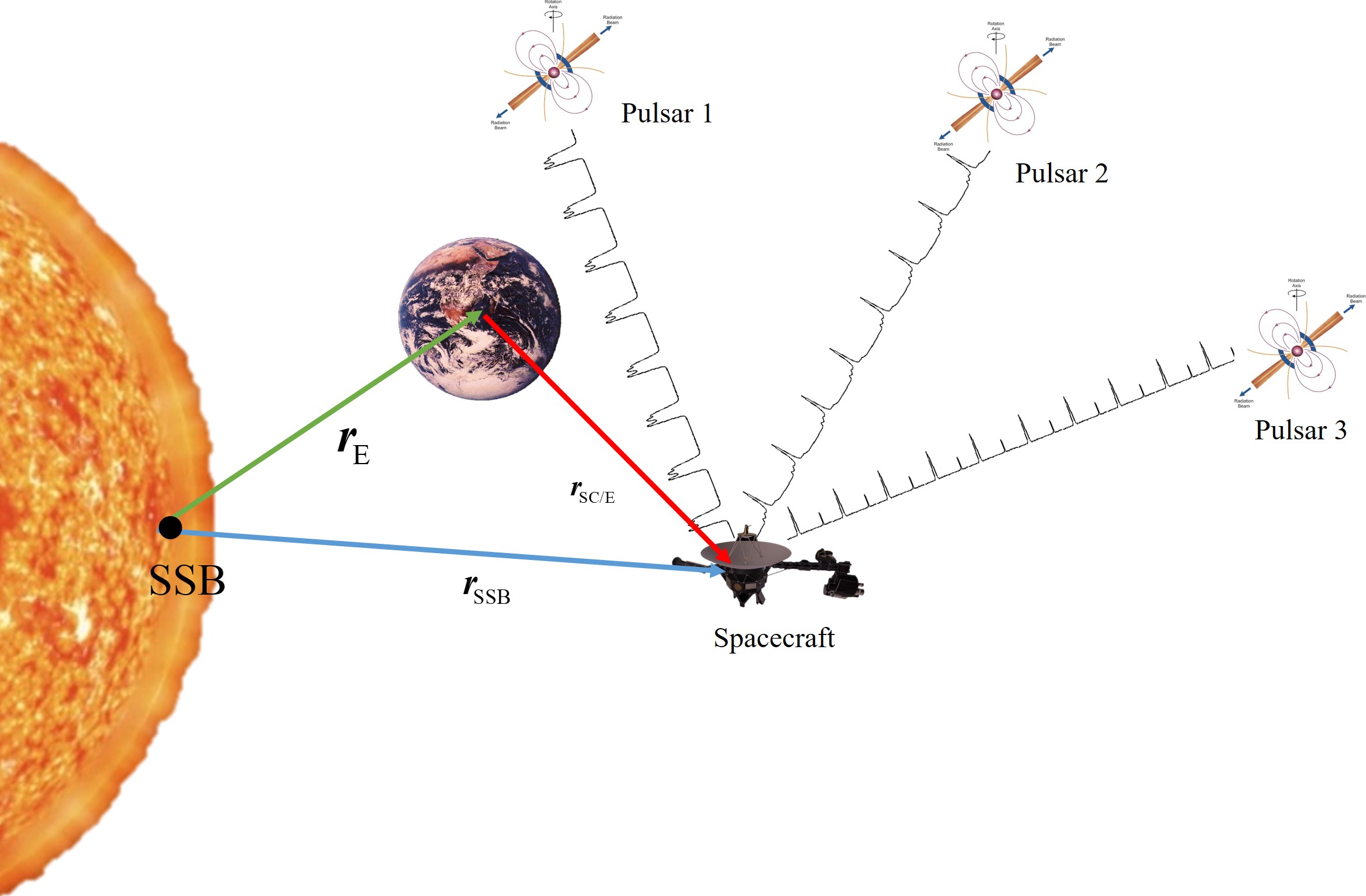}
  \caption{Principle of XNAV for a single spacecraft}\label{fig:principle}
\end{figure}

It can be learned from Eq.(\ref{BC}) that $\boldsymbol{r}_{\mathrm{SSB}}(t)$ projected on the direction of pulsar, $\boldsymbol{n}^{\mathrm{T}}\boldsymbol{r}_{\mathrm{SSB}}(t)$, can be obtained if $t_{\mathrm{SSB}}$ and $t$ are available. When the spacecraft receives signals from three pulsars simultaneously (see Fig.\ref{fig:principle}), $\boldsymbol{r}_{\mathrm{SSB}}(t)$ can be determined via a nonlinear least squares method just like the way to determine the position via GNSS \citep{BUIST2011}. If there are not three pulsars available at the same time, the position of the spacecraft can be estimated by combining the orbit dynamics information with pulsar signals \citep{Zheng2020}.

\subsection{XNAV for multi-spacecraft}
\begin{figure}[!htp]
  \centering
  \includegraphics[width=\columnwidth]{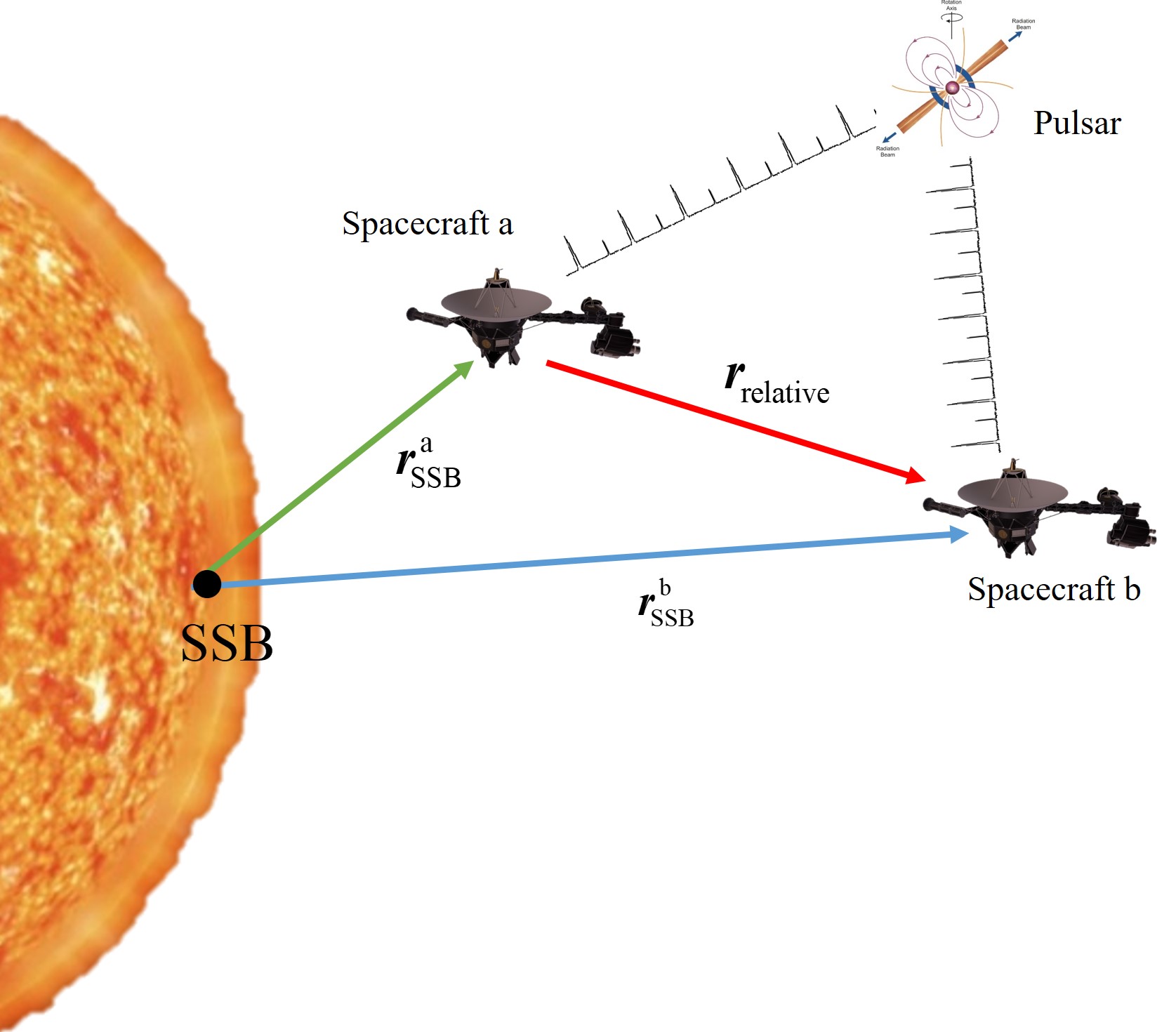}
  \caption{Principle of XNAV for multi-spacecraft}\label{fig:principle_relative}
\end{figure}

Fig.\ref{fig:principle_relative} shows two spacecraft observing one pulsar. Assume $t_{\mathrm{SC}}^{\mathrm{A}}$ and $t_{\mathrm{SC}}^{\mathrm{B}}$ are the pulse TOAs at spacecraft A and spacecraft B, respectively. According to Eq.(\ref{BC}), we have
\begin{equation}
  \label{BC_relative}
  \begin{aligned}
      t_{\mathrm{SSB}}^{\mathrm{A}}-t_{\mathrm{SSB}}^{\mathrm{B}}&=t_{\mathrm{SC}}^{\mathrm{A}}-t_{\mathrm{SC}}^{\mathrm{B}}\\
      &+\frac{1}{c}\boldsymbol{n}^{\mathrm{T}}\left(\boldsymbol{r}_{\mathrm{SSB}}^{\mathrm{A}}(t_{\mathrm{SC}}^{\mathrm{A}})-\boldsymbol{r}_{\mathrm{SSB}}^{\mathrm{B}}(t_{\mathrm{SC}}^{\mathrm{B}})\right)\\
      &+2 \sum_{k} \frac{G M_{k}}{c^{3}} \ln \frac{\boldsymbol{n}^{\mathrm{T}}\boldsymbol{r}_{k}^{\mathrm{A}}(t_{\mathrm{SC}}^{\mathrm{A}})+\left\|\boldsymbol{r}_{k}^{\mathrm{A}}(t_{\mathrm{SC}}^{\mathrm{A}})\right\|}{\boldsymbol{n}^{\mathrm{T}}\boldsymbol{r}_{k}^{\mathrm{B}}(t_{\mathrm{SC}}^{\mathrm{B}})+\left\|\boldsymbol{r}_{k}^{\mathrm{B}}(t_{\mathrm{SC}}^{\mathrm{B}})\right\|}
  \end{aligned}
\end{equation}
where $\boldsymbol{r}_{k}^{\mathrm{A}}$ and $\boldsymbol{r}_{k}^{\mathrm{B}}$ are the positions of the $k$th celestial body relative to the spacecraft A and B, respectively.

In most cases, $\boldsymbol{r}_{k}^{\mathrm{A}} \approx \boldsymbol{r}_{k}^{\mathrm{B}}$ given that the distance between the two spacecraft is far less than the distances between the celestial bodies and the spacecraft. Thus, $\ln \frac{\boldsymbol{n}^{\mathrm{T}}\boldsymbol{r}_{k}^{\mathrm{A}}(t_{\mathrm{SC}}^{\mathrm{A}})+\left\|\boldsymbol{r}_{k}^{\mathrm{A}}(t_{\mathrm{SC}}^{\mathrm{A}})\right\|}{\boldsymbol{n}^{\mathrm{T}}\boldsymbol{r}_{k}^{\mathrm{B}}(t_{\mathrm{SC}}^{\mathrm{B}})+\left\|\boldsymbol{r}_{k}^{\mathrm{B}}(t_{\mathrm{SC}}^{\mathrm{B}})\right\|}$ in Eq.(\ref{BC_relative}) is close to 0. As a consequence, Eq.(\ref{BC_relative}) can be rewritten as
\begin{equation}
  \label{BC_relative2}
  \begin{aligned}
      t_{\mathrm{SSB}}^{\mathrm{A}}-t_{\mathrm{SSB}}^{\mathrm{B}}&=t_{\mathrm{SC}}^{\mathrm{A}}-t_{\mathrm{SC}}^{\mathrm{B}}\\
      &+\frac{1}{c}\boldsymbol{n}^{\mathrm{T}}\left(\boldsymbol{r}_{\mathrm{SSB}}^{\mathrm{A}}(t_{\mathrm{SC}}^{\mathrm{A}})-\boldsymbol{r}_{\mathrm{SSB}}^{\mathrm{B}}(t_{\mathrm{SC}}^{\mathrm{B}})\right)
  \end{aligned}
\end{equation}

As shown in Eq.(\ref{BC_relative2}), assuming $\boldsymbol{r}_{\mathrm{SSB}}^{\mathrm{A}}(t_{\mathrm{SC}}^{\mathrm{A}})-\boldsymbol{r}_{\mathrm{SSB}}^{\mathrm{B}}(t_{\mathrm{SC}}^{\mathrm{B}})=\boldsymbol{r}_{\mathrm{Relative}}$, $\boldsymbol{r}_{\mathrm{SSB}}^{\mathrm{A}}(t_{\mathrm{SC}}^{\mathrm{A}})$, $\boldsymbol{r}_{\mathrm{SSB}}^{\mathrm{B}}(t_{\mathrm{SC}}^{\mathrm{B}})=\boldsymbol{r}_{\mathrm{Relative}}$ and $\boldsymbol{r}_{\mathrm{Relative}}$ can be estimated when three pairs of $t_{\mathrm{SSB}}^{\mathrm{A}}-t_{\mathrm{SSB}}^{\mathrm{B}}$ and $t_{\mathrm{SC}}^{\mathrm{A}}-t_{\mathrm{SC}}^{\mathrm{B}}$ from three pulsars are available simultaneously or sequentially \citep{Emadzadeh2011, KAI2009427}.

\subsection{Techniques supporting XNAV}
As shown in Eqs.(\ref{eq.timing_model}), (\ref{BC}) and (\ref{BC_relative2}), the techniques supporting XNAV are:

\begin{enumerate}[(1)]
  \item Navigation pulsar database:   Pulsars for XNAV are like navigation satellites for GNSS. Unlike the man-made satellites, the location, brightness and modulation properties of pulsars cannot be designed. Thus, pulsars qualified for navigation are rare and their ephemerides need to be updated timely.
  \item X-ray detection system: X-ray detection system for XNAV are like GNSS receivers for GNSS. They detect and record X-ray photons from pulsars.
  \item Pulse TOA estimation: Pulse TOA is an important and basic parameter for XNAV, as illustrated in Eqs.(\ref{BC}) and (\ref{BC_relative2}). However, a spacecraft can only record a series of photons instead of a continuous pulsed signal given that the flux of a pulsar is extremely weak. Moreover, the spacecraft keeps performing an orbit motion, which causes the extraction of pulse TOA from the recorded photons more complex. As a consequence, it is a great challenge to get a reliable and precise pulse TOA.
\end{enumerate}

The advances of the above three techniques will be sequentially introduced in Sections \ref{sect:develp_pdn}-\ref{sect:develop_toa}.

\section{Navigation pulsar database}\label{sect:develp_pdn}
As illustrated in Section \ref{sect:cate}, RPPs are recommended for navigation. Currently, 2200 RPPs have been discovered \citep{Manchester2005}, among which about 150 pulsars emit X-ray radiation \citep{Becker2009}. Approximately 10\% pulsars are MSPs \citep{Becker2013}. MSPs are the most promising reference sources for XNAV because they exhibit very high timing stabilities, which are comparable to the current atomic clocks \citep{Taylor1991, Matsakis1997}.

Table \ref{tbl:pulsar_database} shows the navigation X-ray pulsars used by the past and present missions. The Crab pulsar (or PSR B0531+21) is selected by all missions. Although the Crab pulsar is not a MSP and has glitches in which the spin parameters change suddenly \citep{Shaw2018}, it is much brighter in X-ray than the other pulsars. Therefore, the Crab pulsar can be easily detected and can provide a much more accurate pulse TOA than MSPs when the Crab pulsar and MSPs are observed for the same duration.

The navigation pulsar database contains the following information that describes the timing and location properties of the  selected pulsars:
\begin{enumerate}[(1)]
  \item Position parameters including the declination, the right ascension, and the proper motion of pulsar;
  \item Frequency evolution parameters including the spin frequency and its first- and second-order time derivatives;
  \item The other necessary parameters including the absolute pulse phase for a given epoch, the version of planet ephemerides (NASA's Jet Propulsion Laboratory (JPL) ephemerides for example) , the orbit parameters of binary pulsars, etc.
\end{enumerate}

In practice, the navigation pulsar database is supported by the pulsar timing analysis, which is usually performed by global pulsar timing programs. For example, the North American Nanohertz Observatory for Gravitational Waves (NANOGrav) project \citep{McLaughlin2013} has been timing several pulsars in the pulsar database of SEXTANT program \citep{ray2017}, the Jodrell Bank observes the Crab pulsar \citep{Lyne1993}, and the Parkes Pulsar Timing Array (PPTA) monitors 20 pulsars \citep{Reardon2016}.

\begin{table*}[htbp]
  \caption{X-ray navigation pulsars selected by past and present missions}\label{tbl:pulsar_database}
  \begin{tabular*}{\textwidth}{ll}
  \toprule
   Mission & X-ray pulsars selected for navigation\\ 
  \midrule
   USA experiment\citep{Wood2017} & Crab pulsar\\
   \multirow{2}{1in}{XNAV program\citep{Sheikh2006}} & PSR B1937+21, PSR B1957+20, SAX J1808.4-3658, PSR B1821-24, XTE J1751-305, XTE J1807-294\\
    &Crab pulsar, PSR B0540-69\\
   \multirow{2}{1in}{SEXTANT program\citep{ray2017}} & PSR B1937+21, PSR B1821-24, PSR J0218+4232, PSR J0030+0451, PSR J1012+5307, PSR J0437-4715\\
   &PSR J2124-3358, PSR J0751+1807, PSR J1024-0719, PSR J2214+3000, Crab pulsar \\
   XPNAV-1\citep{Huang2019}& Crab pulsar\\
   Insight-HXMT\citep{Zheng_2019}& Crab pulsar\\
  \bottomrule
  \end{tabular*}
  \end{table*}

\section{X-ray detection system}\label{sect:develp_dect}
The pulse X-ray emission of a RPP is usually non-thermal, which has a power law spectrum with a photon index of about $-2$, i.e., the flux of the pulse photon in the soft X-ray band (photon energy less than 10 kilo electron Volt (keV)) is much higher than that in the hard X-ray band (energy above 10 keV) \citep{Sheikh2006,Kalemci2018}. More photons indicate a more accurate pulse TOA estimation (see Section \ref{sect:develop_toa}). Therefore, the soft X-ray radiation from pulsars is recommended for navigation. The background, which comes from the diffuse Cosmic X-ray Background (CXB), space particles and the non-pulse portion of the pulsar, is commonly detected along with the X-rays of pulsars \citep{Kalemci2018}. As a consequence, an X-ray detection system consists of an optics system, designed to focus X-ray photons, and an X-ray detector counting X-ray photons collected by the optics system.

\subsection{Optics system}
Two broad categories of optics systems  are used in XNAV observations: collimators and focusing X-ray telescopes. A collimator restricts the Field of View (FOV) of the whole detection system. In this way, only the background within a narrow portion of the space can enter the detection system. On the other hand, a focusing X-ray telescope focuses the source flux onto the X-ray detector in the focal plane \citep{Ramsey1993}. As will be seen in Section \ref{set:gio}, only the background arriving within a small incidence angle can be collected. Therefore, both the collimator and focusing X-ray telescope can effectively reduce the background.

\subsubsection{Collimators}
\begin{figure}[!h]
  \centering
  \includegraphics[width=\columnwidth]{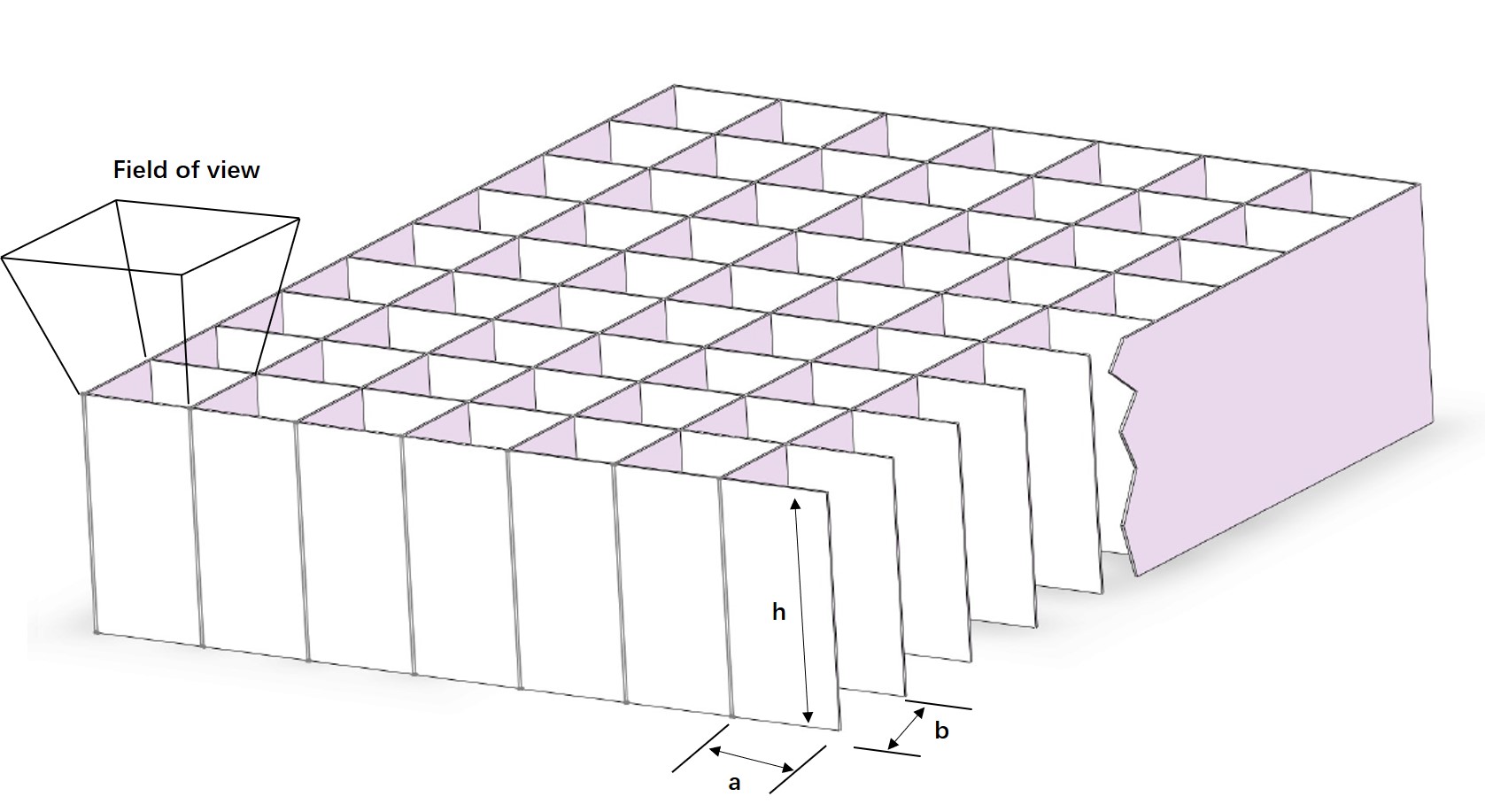}
  \caption{A simple slat collimator \citep{Ramsey1993}}\label{fig:collimator}
\end{figure}

Fig.\ref{fig:collimator}\citep{Ramsey1993} shows a simple slat collimator with an FOV of $\mathrm{arctan}\left(a/h\right)\times \mathrm{arctan}\left(b/h\right)$ where $a$, $b$ and $h$ are the length, width, and height of a slat respectively. All photons or particles arriving at angles larger than $\mathrm{arctan}\left(a/h\right)$ and $\mathrm{arctan}\left(b/h\right)$ are attenuated by the walls of collimator structure. In other words, the photons or particles from the sources with angular positions relative to the targeted source outside the FOV are rejected. Thus, for collimators, a small FOV indicates a strong capability of rejecting background.

Because of their simple structures, collimators were used in a number of X-ray missions including Uhuru, the first X-ray satellite launched in 1970 \citep{Fabian1975} (see \textcolor{red}{Table \ref{tbl:satellite_collimator}\citep{Fabian1975, Giacconi1971, Sanford1974,Gursky1978,Rothschild1979,Turner1981,Turner1989,Jahoda2006,ZhangSN2020,Smith1976}}). Collimators are also used in X-ray imaging, with scanning observations and image reconstruction techniques \citep{bradt1968modulation}.

\begin{sidewaystable*}[!htbp]
  \caption{Past and active X-ray telescopes employing collimators}\label{tbl:satellite_collimator}
  \resizebox{\textwidth}{!}{
  \begin{tabular}{ccccccc}
  \toprule
  \multirow{2}{*}{Spacecraft} & \multirow{2}{*}{Year} & \multirow{2}{*}{Instrument} &
  \multicolumn{2}{c}{Optics system}& \multirow{2}{*}{Detector}& \multirow{2}{*}{Total effective area ($\mathrm{cm^{2}}$)}\\
  \cline{4-5}
   & & & FOV &  Energy range (keV)& &\\
    \midrule
  Uhuru \citep{Fabian1975, Giacconi1971}  & $1970-1973$&- &$0.52^{\circ}\times 0.52^{\circ}$ and $5.2^{\circ}\times 5.2^{\circ}$ & 2-20& Proportional counter&1680\\
  \multirow{2}{*}{Ariel V\citep{Smith1976}}  & \multirow{2}{*}{$1974-1980$}& LE& $0.75^{\circ}\times 10.6^{\circ}$& 1.2-5.8 & Proportional counter& 145\\
  & & HE & $0.75^{\circ}\times 10.6^{\circ}$&2.4-19.8& Proportional counter& 145\\
  OAO-3\citep{Sanford1974} & $1972-1981$&-&$2.5^{\circ}\times 3.5^{\circ}$& 2.5-10&Proportional counter &  17.8\\
  \multirow{5}{*}{HEAO-1\citep{Gursky1978,Rothschild1979}} & \multirow{5}{*}{$1977-1979$}&LASS& varied between $1^{\circ}\times 4^{\circ}$ and $1^{\circ}\times 0.5^{\circ}$& 0.25-25& Proportional counter & $7\times (1350-1900)$\\
  & & \multirow{3}{*}{CEX} &  &0.15-3.0 & Proportional counter & $2\times 400 $\\
  & &  &\multirow{2}{*}{various setting: $1.5^{\circ}\times 3^{\circ}$, $3^{\circ}\times 3^{\circ}$, $3^{\circ}\times 6^{\circ}$} &1.5-20 & Proportional counter & 800\\
  & &  &  &2.5-60 & Proportional counter & $3 \times 800$\\
  & & \multirow{2}{*}{MC} & \multirow{2}{*}{$4^{\circ}\times 4^{\circ}$} & \multirow{2}{*}{0.9-13.3}& Proportional counter & $400$\\
  & &  &  & & Proportional counter & $300$\\
  EXOSAT\citep{Turner1981}  & $1983-1986$&ME &$45'\times 45'$& 1.3-15, 5-55 & Proportional counter & $1800$\\
  Ginga\citep{Turner1989}  & $1987-1991$&LAC &$0.8^{\circ}\times 0.7^{\circ}$ &1.5-30 & Proportional counter & $4000$\\
  RXTE\citep{Jahoda2006}  &  $1995-2012$&PCA  &$1^{\circ}$&2-60&Proportional counter& $6500 $\\
  \multirow{3}{*}{Insight-HMXT\citep{ZhangSN2020}}  &  \multirow{3}{*}{$2017-$} &LE & $15 \times(1.1^{\circ}\times 5.7^{\circ})$ and $2 \times(5.7^{\circ}\times 5.7^{\circ})$& 1-12&SCD & $384$\\
  & &ME& $16\times (1^{\circ}\times 4^{\circ})$ and $2\times (4^{\circ}\times 4^{\circ})$& 8-35&Si-PIN& $952$\\
  & &HE& $16\times (1.1^{\circ}\times 5.7^{\circ})$ and $2\times 5.7^{\circ}\times 5.7^{\circ}$&20-350&NaI(CsI)& $5100$\\
  \bottomrule
\end{tabular}}
  
1) Note: HEAO-1 $-$ High Energy Astronomy Observatory-1; EXOSAT $-$ European X-ray Observatory SATellite; RXTE $-$ Rossi X-ray Timing Explorer; LE $-$ Low Energy system, HE $-$ High Energy system; LASS $-$ Large Area Sky Survey experiment; LAC $-$ Large Area Proportional Counter; CXE $-$ Cosmic X-ray Experiment; PCA $-$ Proportional Counter Array
\end{sidewaystable*}

\subsubsection{Focusing X-ray telescopes}\label{set:gio}
X-rays can be focused with the grazing incidence optics. The mirrors comprising focusing X-ray telescopes are typically multi-shell like structures coated with a heavy metal, such as gold or nickel \citep{Smith2006}. Soft X-ray photons can be reflected from the surface of mirror, when their angles of incidence with respect to the surface are less than \citep{Peter2022}
\begin{equation}
  \theta_{\mathrm{c}}=37 \frac{\rho^{1 / 2}}{E}
\end{equation}
where $\rho$ is the density of coating material and $E$ is the energy of photon.

The reflectivity of gold is a function of energy and angle of incidence\citep{Peter2022}, which increases as the energy and angle of incidence reduce. In practice, the angle of incidence is typically less than $1^{\circ}$. However, a low angle of incidence can be achieved at the cost of a long focal length of the telescope.

In order to achieve a good reflectivity and good imaging quality, focusing X-ray telescopes require much precise figuring and ultra-high-quality mirrors with roughness better than 1 nm \citep{Ramsey1993}. The above requirements all indicate that the fabrication of focusing X-ray telescopes is expensive and labor-intensive.

For a focusing X-ray telescope, its effective area depends on the energies of incoming photons because the reflectivity is a function of photons' energies and thus is with the unit of "$\mathrm{cm}^{2}@b\mathrm{keV}$", where $@b\mathrm{keV}$ means "at $b$ $\mathrm{keV}$". Besides this, the angular resolution quantified by the Half-Power Diameter (HPD) is also an important parameter to describe the imaging capability of a focusing telescope\citep{Dell2010}. HPD is defined as the angular diameter in the focal plane, which contains half the flux (at a given energy) focused by the telescope.

Although there are various geometries that can achieve grazing incidence optics \citep{Gorenstein2012}, the Wolter-I geometry is most frequently used in X-ray telescopes (see \textcolor{red}{Table \ref{tbl:wolter}\citep{Tousey1977,Giacconi1979,Taylor1981SSR,Korte1981,Aschenbach1988, Dell2010,Serlemitsos1995,Gorenstein2010, Smith2006,David2012,Burrows2005,Serlemitsos2007,Harrison2013,Singh2014,Takahashi2018, Toshiki2016,LI2018,Okajima2016,Friedrich2008,Pavlinsky2021}}). Since the FOV of a Wolter-I geometry is small, recently, a lobster-eye geometry is used to achieve wide field X-ray imaging. A lobster-eye geometry can observe simultaneously multiple sources locating relative far away, such as $10^{\circ}$ from each other, and can monitor transient X-ray sources. This section will focus on the two geometries.

(1) Wolter-I geometry

\begin{figure}[h]
  \centering
  \includegraphics[width=\columnwidth]{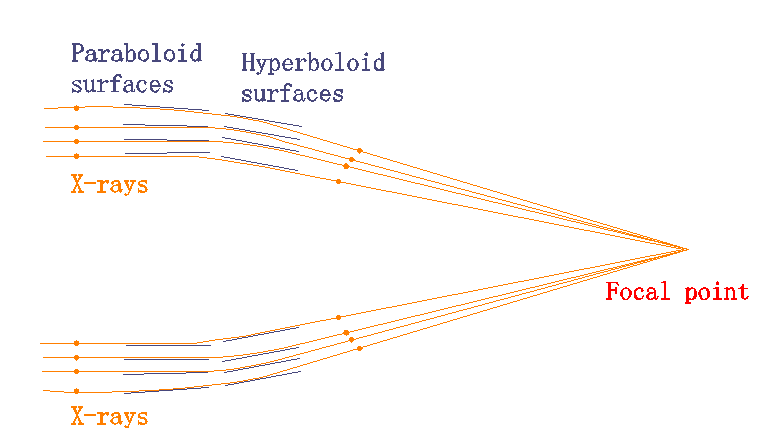}
  \caption{Principle of Wolter-I geometry\citep{Wolter1952}}\label{fig:wolter-I}
\end{figure}

\begin{figure}[!h]
  \centering
  \includegraphics[width=\columnwidth]{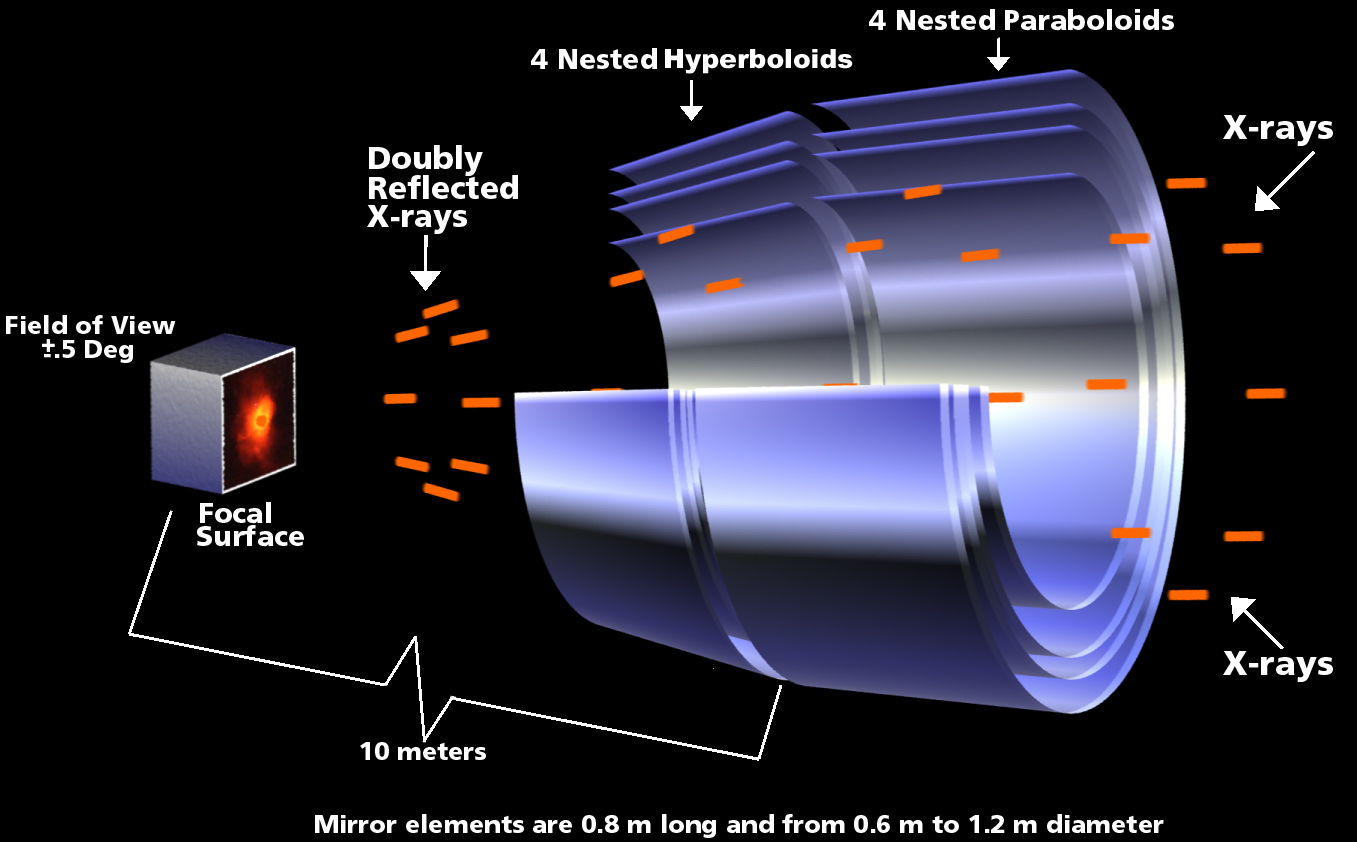}
  \caption{Nested X-ray mirrors on Chandra X-ray Observatory (Image credit: NASA, Chandra, and the Smithsonian Astrophysical Observatory.)}\label{fig:cxcmirror}
\end{figure}

Almost the focusing X-ray telescopes that have been in orbit were fabricated in accordance with or approximating the Wolter-I geometry \citep{Gorenstein2010} with the principle shown in Fig.\ref{fig:wolter-I} \citep{Wolter1952}. The incoming rays experience two grazing incidence reflections from two confocal paraboloidal and hyperboloidal mirrors sections. Several pairs of paraboloidal and hyperboloidal mirrors can be nested to increase the effective area. Fig.\ref{fig:cxcmirror} shows the nested mirrors on the Chandra X-ray Observatory.

\begin{sidewaystable*}[!htbp]
	\caption{Past and active X-ray telescopes using Wolter-I geometry}\label{tbl:wolter}
  \resizebox{\textwidth}{!}{
	\begin{tabular}{clccccccccc}
  \toprule
	\multirow{3}{*}{Spacecraft} & \multirow{3}{*}{Year} & & &Optics system& & & &  & &   \\
  \cline{3-8}
   & & Instrument &  Energy range & Angular resolution/HPD & Focal length & Reflecting 	& Number of* & Detector & Total effective area\\
    & &  &  (keV) &  &  (m) &surface &nested shells & on the focal plane&  \\
  \midrule
  Skylab\citep{Tousey1977} & 1973$-$1974 & ATM & 0.23-4.1 & 3" &  1.903 & Fused silica& 1& Proportional counter&9 $\mathrm{cm}^{2}@1\mathrm{keV}$\\
  HEAO-2 (Einstein)\citep{Giacconi1979}& 1978$-$1981 &    & 0.2-4 & 4" & 3.4 & quartz & 4 &Proportional counter&$ 400\mathrm{cm}^{2}@0.25$ keV \\
  EXOSAT\citep{Taylor1981SSR,Korte1981}& 1983$-$1986 &  LE  & 0.05-2 & 24" & 0.28 & Gold & 4& Proportional counter&$38\mathrm{cm}^{2}@1.5$keV\\
  ROSAT\citep{Aschenbach1988, Dell2010} &1990$-$1999 &  & 0.5-1.5 & 5" &2.4 & Gold & 4 & Proportional counter& $400\mathrm{cm}^{2}@1$keV \\
  \multirow{2}{*}{ASCA\citep{Serlemitsos1995}} &\multirow{2}{*}{1993$-$2001} &  & \multirow{2}{*}{0.4-10} & \multirow{2}{*}{2.9$'$} & \multirow{2}{*}{3.5}&\multirow{2}{*}{Gold} & \multirow{2}{*}{$4\times 30$}& Proportional Counter & \multirow{2}{*}{$4 \times 300\mathrm{cm}^{2}@1$keV} \\
   & &  &  & & & & & CCD &  \\
  Chandra\citep{Gorenstein2010, Smith2006,Dell2010} &1999$-$ &  & 0.1-10 & 0.5" & 10.0& Iridium& 4& CCD &$800\mathrm{cm}^{2}@0.25$keV \\
  XMM-Newton\citep{David2012} &1999$-$  &  & 0.1-12 & 15" &7.5 & Gold & $3\times 58$& CCD& $3 \times 1500\mathrm{cm}^{2}@1$ keV \\
  \textit{Swift}\citep{Burrows2005} & 2004$-$ &  & 0.2-10 & 15" &3.5& Gold & 12& CCD&$110\mathrm{cm}^{2}@1.5$ keV \\
  \multirow{2}{*}{Suzaku\citep{Serlemitsos2007}} & \multirow{2}{*}{2005-} & XRT-I & \multirow{2}{*}{0-10} & \multirow{2}{*}{2$'$}& 4.75&\multirow{2}{*}{Gold}&175 &CCD &$450\mathrm{cm}^{2}@1.5$keV  \\
  &  & XRT-S &  & &4.5 & &168&X-Ray Spectrometer & $450\mathrm{cm}^{2}@1.5$keV \\
  \multirow{2}{*}{\textit{NuSTAR}\citep{Harrison2013}} & \multirow{2}{*}{2012$-$} &  & \multirow{2}{*}{3-78.4} & \multirow{2}{*}{58"} & \multirow{2}{*}{10.14} & Platinum/Carbon (inner surface)& \multirow{2}{*}{133}& \multirow{2}{*}{CdZnTe detector} &\multirow{2}{*}{$847\mathrm{cm}^{2}@9$keV} \\
  & & & & & & Tungsten/Silicon (outer surface)& & & \\
  ASTROSAT\citep{Singh2014} & 2015$-$ & SXT & 0.3-8 & 2$'$ &2 & Gold& 40& CCD&$128\mathrm{cm}^{2}@1.5$keV  \\
  Hitomi\citep{Takahashi2018, Toshiki2016} & 2016$-$2016& SXI &0.4-12 & 1.3$'$ & 5.6 & Gold& 203&CCD& $ 370\mathrm{cm}^{2}@1$keV \\
  XPNAV-1\citep{LI2018} & 2016$-$ & iFXPT &0.5-10 & 15$'$ & 1.15 & Gold& 4&SDD& $ 2.8\mathrm{cm}^{2}@1$keV \\
  ISS\citep{Okajima2016} & 2017$-$ & NICER & 0.2-12 &  & 1.085 & Gold& $56\times 24$&SDD& $ 1793\mathrm{cm}^{2}@1.5$keV\\
  \multirow{2}{*}{SRG\citep{Friedrich2008,Pavlinsky2021}} & \multirow{2}{*}{2019$-$} &  ART-XC & 4-30 & 30-35" & 2.7& Iridium & 28 & CdZnTe detector&$385\mathrm{cm}^{2}@8.1$keV\\
  & & eROSITA & 0.2-10 & 15" & 1.6& Gold& $7\times 54$ & CCD&$7\times 300\mathrm{cm}^{2}@1$keV\\
  \bottomrule
	\end{tabular}}

  1) Note: ATM $-$ Apollo Telescope Mount; HEAO-2 $-$ High Energy Astronomy Observatory-2; ROSAT $-$ ROentgen SATellite; ASCA $-$ Advanced Satellite for Cosmology and Astrophysics; Chandra $-$ Chandra X-ray Observatory; XMM-Newton $-$ X-ray Multi-Mirror-Newton; XRT-I $-$ X-ray telescope for X-ray Imaging Spectrometer; XRT-S $-$ X-ray telescope for X-ray Spectrometer; ASTROSAT $-$ Astronomy Satellite; SXT $-$ Soft X-ray Telescope; SXI $-$ Soft X-ray camera; iFXPT $-$ incidence Focusing X-ray Pulsar Telescope; SRG $-$ Spectrum Roentgen Gamma; ART-XC $-$ Astronomical Roentgen Telescope–X-ray Concentrator; eROSITA $-$ extended ROentgen Survey with an Imaging Telescope Array

  *) A shell denote a pair of paraboloidal and hyperboloidal mirrors.

  $\dagger$) NICER employs only paraboloidal mirrors and cannot image.
\end{sidewaystable*}

(2) Lobster-eye optics

\begin{figure}[!h]
  \centering
  \includegraphics[width=\columnwidth]{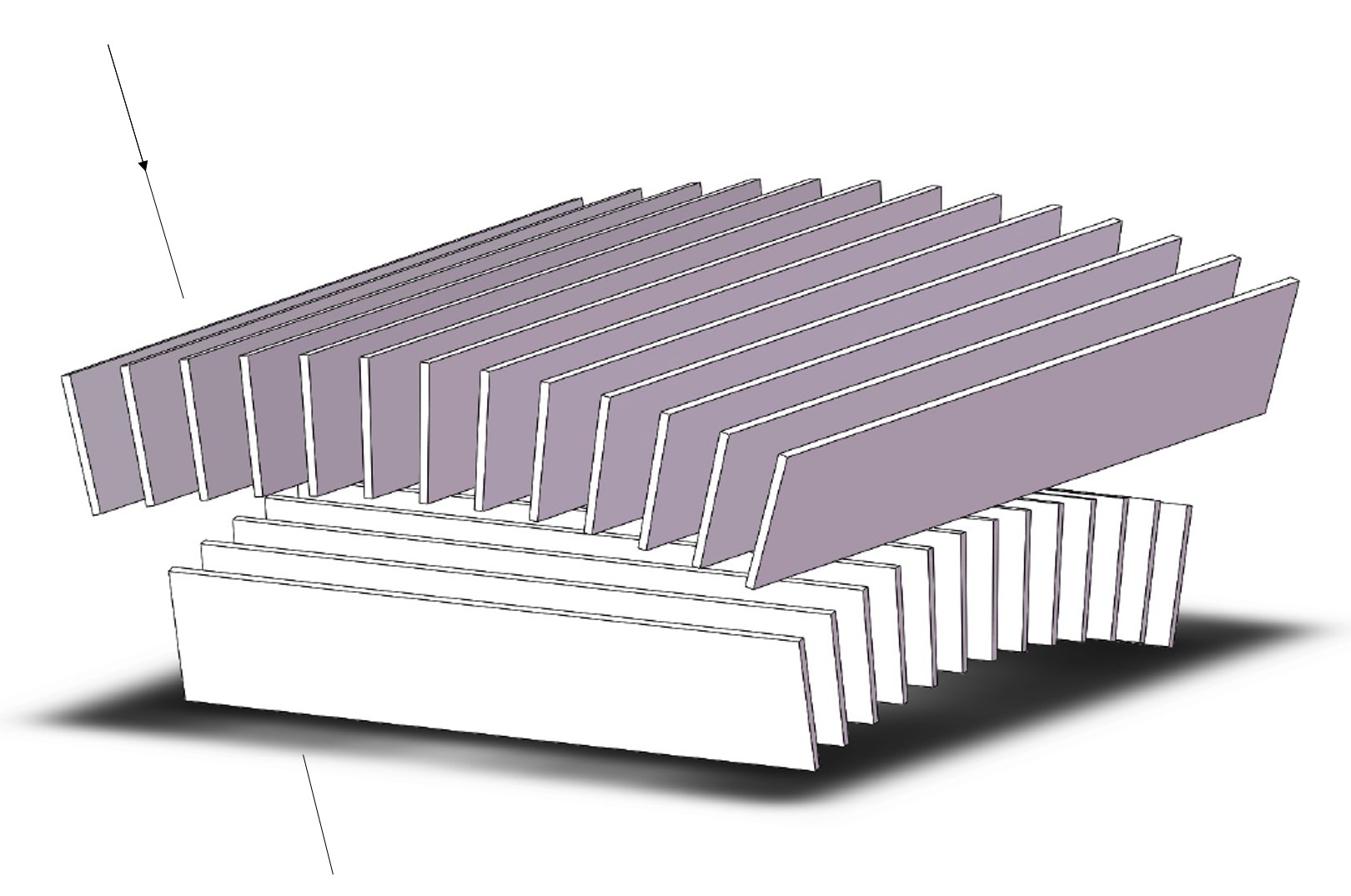}
  \caption{Schmidt arrangement for a lobster-eye geometry}\label{fig:schmidt}
\end{figure}

\begin{figure}[!h]
  \centering
  \includegraphics[width=\columnwidth]{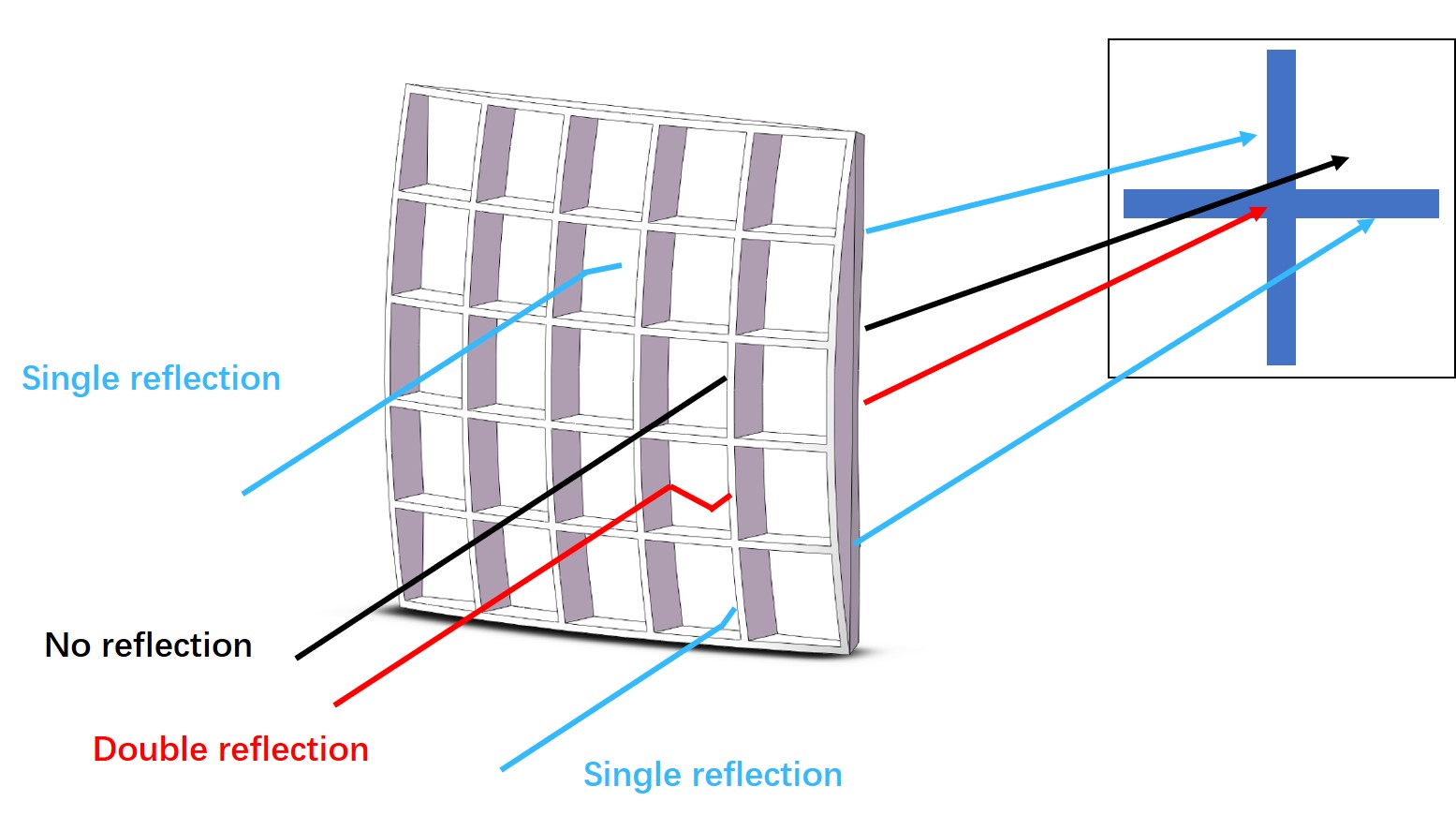}
  \caption{Angel arrangement for a lobster-eye geometry}\label{fig:angel}
\end{figure}

This optics mimics the eyes of lobsters and was suggested independently by Schmidt \citep{Schmidt1975} in 1975 and by Angel \citep{Angel1979} in 1979. The Schmidt arrangement employs the orthogonal sets of mirrors (see Fig.\ref{fig:schmidt}) and the Angel arrangement utilizes array of reflecting cells (see Fig.\ref{fig:angel}). The lobster-eye optics has a wide FOV, up to maximum of $2\pi$ \citep{Pina2019}. This feature makes the lobster-eye optics more suitable for the all-sky survey than the Wolter-I geometry, the FOV of which is typically less than $1^{\circ}$ \citep{Gorenstein1987, Fraser2002}.

The lobster-eye optics can be constructed as a \\
one-Dimensional (1D) or two-Dimensional (2D) system. The Schmidt arrangement shown in Fig.\ref{fig:schmidt} is a 2D system, which consists of two orthogonal 1D systems. The Angel arrangement is also a 2D system. The incoming photons focused to form a point with tails forming a cross like image, in which the photons at the center point are reflected twice and those in the wings are only reflected once \citep{Yuan2018}. It appears that the Schmidt arrangement can practically achieve an angular resolution of the order of one tenth of a degree or better, and the angular resolution of the Angel arrangement can be as good as a few arc seconds, if very precise reflecting cells can be fabricated \citep{Pina2019}.

Nowadays, both the Schmidt and Angel arrangements are under development. Ref.\citenum{Pina2019} designs an X-ray optics system for CubSat demonstrator based on the Schimdt arrangement. Ref.\citenum{George2002} uses the microchannel plate technology to fabricate the reflecting cells in the Angel arrangement. Ref.\citenum{Yuan2018EP} employs the micro-pore optics to realize the Angel arrangement.

In the near future, the Einstein Probe (EP) \citep{Yuan2018SSPMA} will employ the lobster-eye optics to form a large FOV monitor to detect X-ray transients and gamma-ray bursts. In order to test the lobster-eye optics for EP, the EP-WXT Pathfinder, an experimental module for the Wide-field X-ray Telescope (WXT) of EP, was loaded on the satellite SATech-01 launched on July 27, 2022. The EP-WXT Pathfinder released on its first results on August 27, 2022. The Space-based multi-band astronomical Variable Objects Monitor (SVOM) \citep{tz2009,Feldman2017} will employ a narrow-field-optimised lobster-eye telescope to promptly detect and accurately locate gamma-ray bursts afterglows. These two satellites will be launched in the next couple of years.

\subsection{Detectors}
X-ray detectors aim to turn the collected X-rays into electronic signals and further into data that can be stored for later analysis \citep{arnaud_smith_siemiginowska_2011}. These data typically include the arrival time, energy, and the location on the detector of each photon detected. The commonly used X-ray detectors are proportional counters, Charge-Coupled Devices (CCDs), Silicon Drift Detectors (SDDs), scintillation detectors, CdZnTe detectors, etc.

\subsubsection{Proportional counters}
Proportional counters were stemmed from gases detectors, which were first applied to detect cosmic rays and other particles in laboratory \citep{xu1987}, and have been used in soft X-ray astronomy for nearly six decades.

\begin{figure}[!h]
  \centering
  \includegraphics[width=\columnwidth]{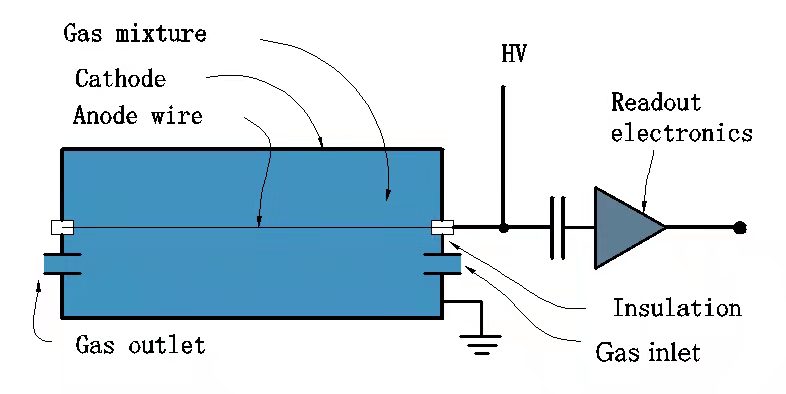}
  \caption{Schematic diagram of a proportional counter}\label{fig:gpc}
\end{figure}

A standard proportional counter is a gas-filled chamber with a coaxial thin wire as shown in Fig.\ref{fig:gpc}. The electrical field is generated by applying positive High Voltage (HV) to the anode \citep{Winkler2015}. X-rays entering the chamber ionize the gas, producing electrons. The electrons can be accelerated in the electric field and further ionize gas atoms by collisions and cause charge multiplication. In proportional counter, the HV is tuned to maintain a linear relation between the input photon energy and the charge \citep{arnaud_smith_siemiginowska_2011}. This is the very reason for its name.

As shown in Tables \ref{tbl:satellite_collimator} and \ref{tbl:wolter}, the combinations of proportional counters and collimators were widely applied to X-ray astronomy satellites, especially those launched before 2000. It is because, the observation of the weak photon fluxes from cosmic X-ray sources with non-imaging detection systems (detectors with collimators) requires large-area detectors, which can be realized by multi-anode multilayer proportional counters under the level of fabrication then.

In order to improve the energy resolution of the proportional counter, the Gas Scintillation Proportional Counter (GSPC) was proposed \citep{CONDE19677} and used aboard the EXOSAT \citep{White1990}. However, GSPCs have now been superseded by CCDs (see Section \ref{sect:ccd}) for soft X-ray band and CdZnTe detectors (see Section \ref{sect:czt}) for hard X-ray band \citep{arnaud_smith_siemiginowska_2011}.

Proportional counters may suffer from an aging problem because of the cracking of gas molecules during normal operation \citep{Tr2008}. The USA experiment (see Section \ref{sect:usa}) was aborted because of the leakage of gases filled in the proportional counters.

\subsubsection{CCD}\label{sect:ccd}

The CCD was invented at Bell Laboratories in 1969, and was first employed as the focal plane detector for X-ray imaging by the ASCA satellite in 1993. Since then, CCDs are the most widely used detectors in X-ray astronomy because a CCD detector usually has small pixel size, large chip size, good energy resolution, and low noise \citep{Valentina2020}.

\begin{figure}[!h]
  \centering
  \includegraphics[width=\columnwidth]{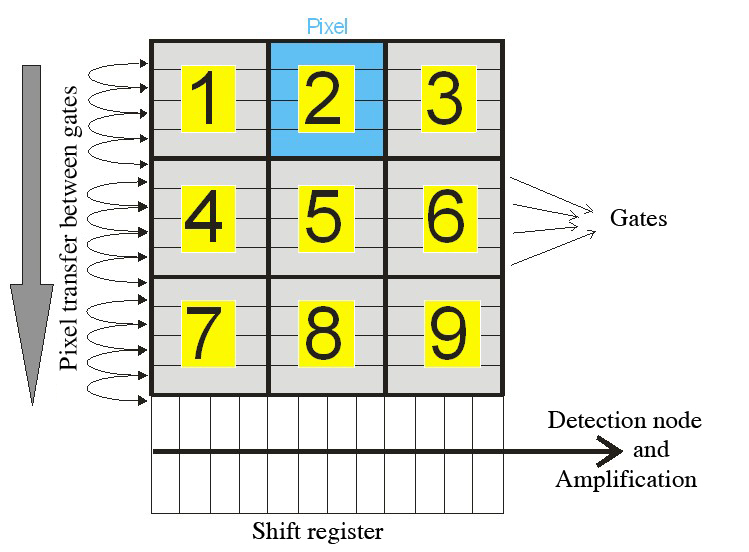}
  \caption{Electrons transfer principle for a $3\times 3$ pixels CCD\citep{CCD_html}}\label{fig:ccd}
\end{figure}

A conventional CCD is an array of small and thin electrodes of metal-oxide-semiconductor capacitors. The incoming X-rays interact with the semiconductor substrate, producing electron-hole pairs. An electric field can be produced, when the gates between neighboring pixels have potentials \citep{Valentina2020}. In this way, the electrons can be stored in pixels, and can be transferred, pixel to pixel, from the semiconductor to a readout amplifier, by tuning the gate potentials (see Fig.\ref{fig:ccd}).

\begin{figure}[!h]
  \centering
  \includegraphics[width=\columnwidth]{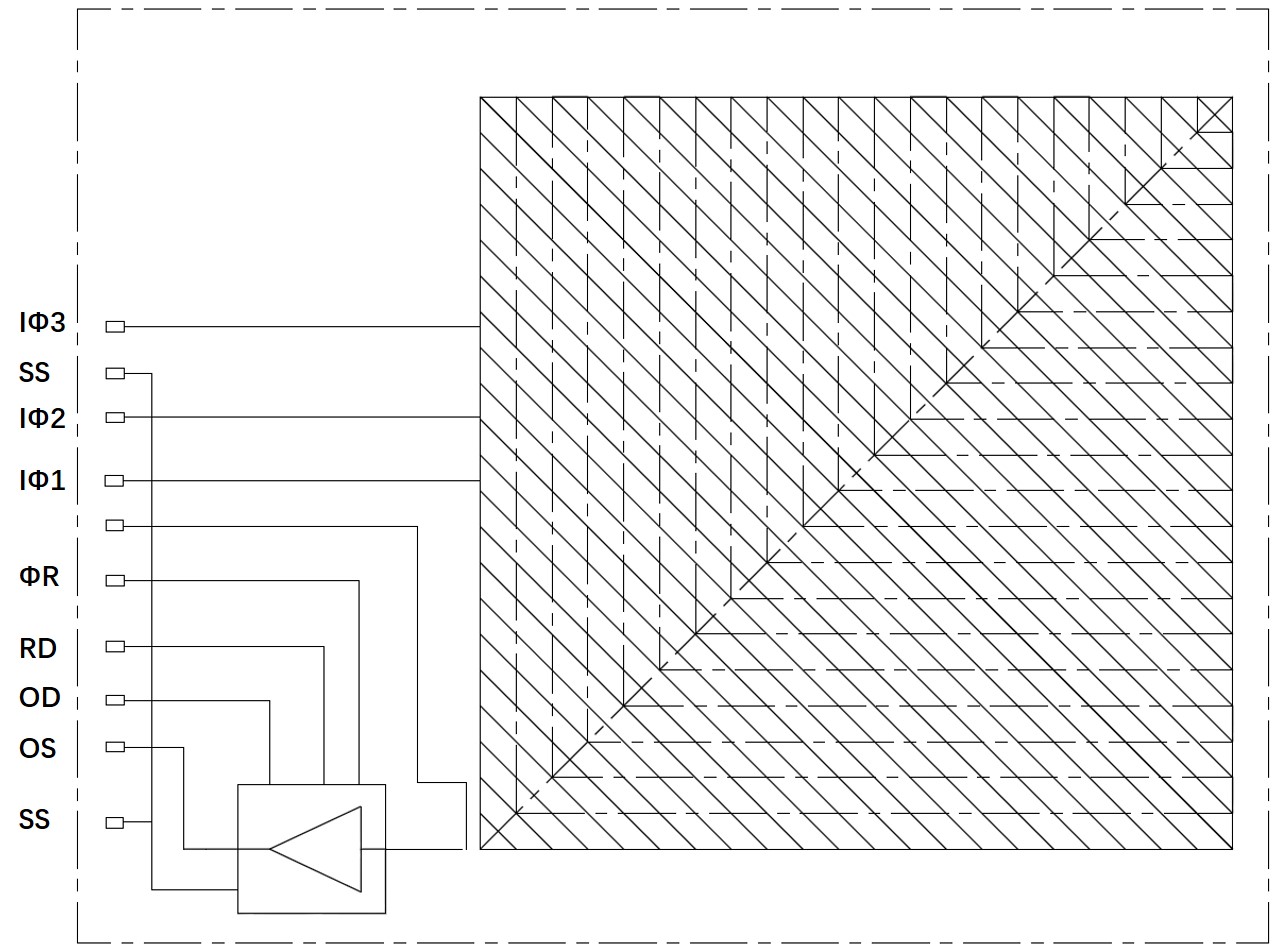}
  \caption{Schematic diagram of SCD}\label{fig:scd}
\end{figure}

Although CCDs have a high energy resolution up to about 130 eV@ 6keV and a spatial resolution on the order of 10 ${\rm\mu}$m, their read-out typically cost several seconds \citep{Nakajima2009}. In order to accelerate the read-out, two variations on CCD have been designed, including the pnCCD \citep{Ihle_2017} and the Swept Charge Devices (SCD) \citep{Gow2012}. In a pnCCD, the pixels are aligned as column parallel, and each column has a dedicated readout channel \citep{Andritschke2008, Meidinger2006}. In an SCD, the electrons generated by interactions between X-rays and the semiconductor substrate are simultaneously or consecutively swept towards the central diagonal channel, and then are transferred to the single read-out node located in one corner of the devices (see Fig.\ref{fig:scd}). In this way, an SCD can improve the time resolution at the cost of sacrificing imaging capability \citep{LOWE2001568}. Up to now, pnCCD has been applied to the XMM-Newton and eROSITA, and will operate on the SVOM and EP. SCDs have been applied to the Insight-HXMT.

\subsubsection{SDD}\label{set:sdd}
\begin{figure}[!h]
  \centering
  \includegraphics[width=\columnwidth]{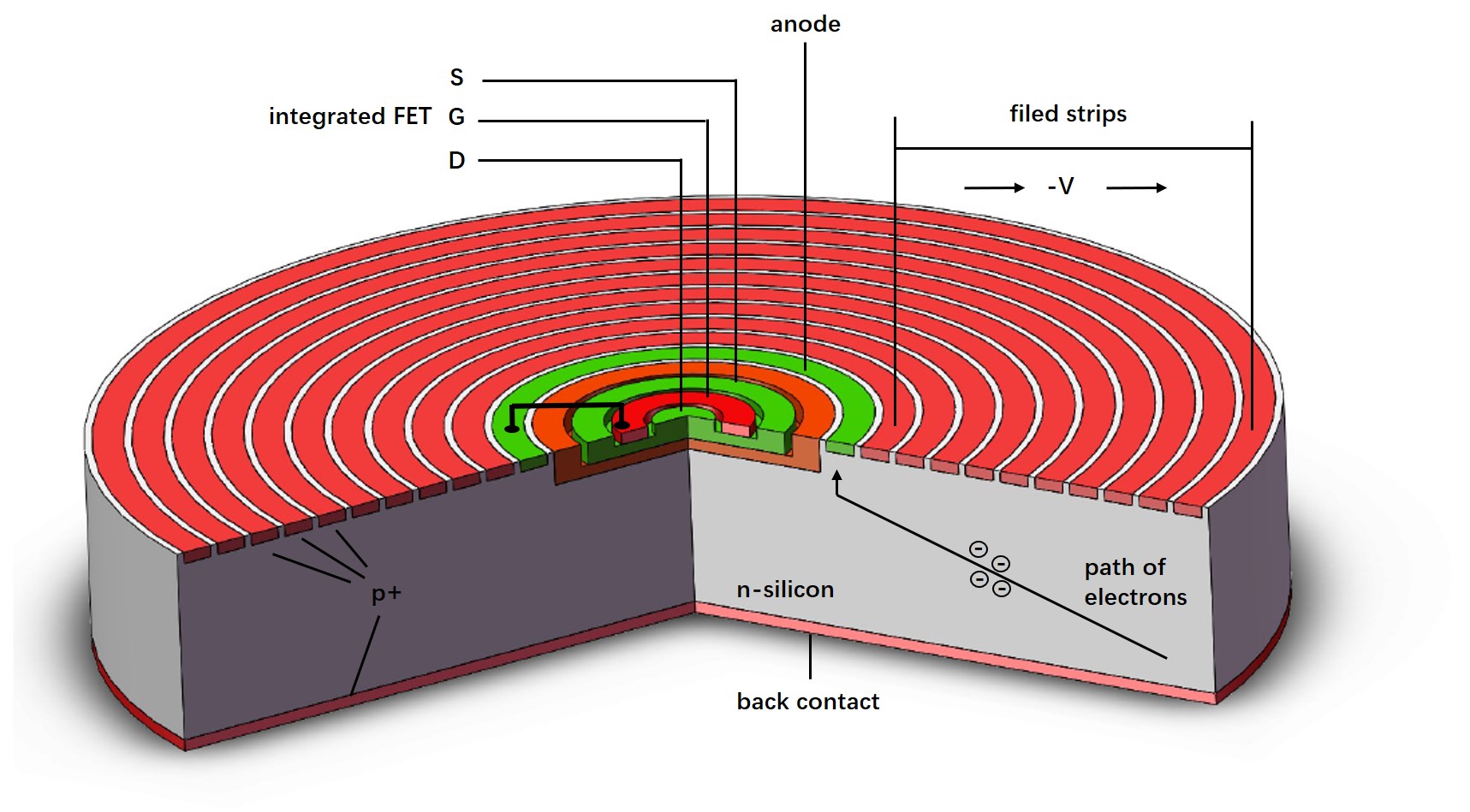}
  \caption{Schematic diagram of SDD}\label{fig:sdd}
\end{figure}

SDDs were first proposed in 1983 \citep{Gatti1984}. An SDD consists of a radial electric field, which terminates in a very small collecting anode on one face of the device (see Fig.\ref{fig:sdd}). Electrons generated by the interactions between X-rays and the semiconductor are guided along these electric field lines to the anode. On the other hand, a small anode indicates a low device capacitance, and finally produces a small electronic noise \citep{Takahashi2001}. Electrons move fast in the electric field, and so can be collected in a very short time. SDDs thus have a very high time resolution. However, one SDD pixel needs one read-out channel, which makes it almost impossible to obtain high spatial resolution images like CCDs.

SDDs have been successfully applied to NICER and XNAV-1, and are planned to be employed by CubeX, XNavSat and the enhanced X-ray Timing and Polarimetry (eXTP) \citep{Zhang2016SPIE}.

\subsubsection{Scintillation detector}
\begin{figure}[!h]
  \centering
  \includegraphics[width=\columnwidth]{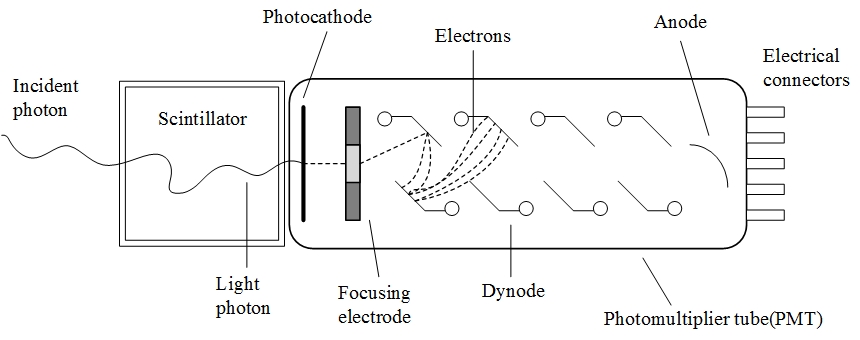}
  \caption{Schematic diagram of a scintillation detector comprising a scintillation material coupled to a PMT}\label{fig:sd}
\end{figure}

Scintillation detectors are employed to detect X-rays with energies higher than several keV. A scintillation detector is a combination of a crystal, which emits optical light photons when absorbing X-ray photons, and a light sensor, which collects the optical photons \citep{Valentina2020} (see Fig.\ref{fig:sd}). The light sensors are typically Photo-Multiplier Tubes (PMT), photodiodes, silicon photo-multipliers, and SDD. In practice, a phoswich, a combination of two different crystals, is usually employed instead of a single crystal. The NaI(Tl)/CsI(Na) phoswich has been applied to RXTE \citep{Rothschild_1998} and Insight-HXMT \citep{Liu2020SCPMA}.

\subsubsection{CdZnTe detector}\label{sect:czt}
Cadmium Telluride (CdTe) and Cadmium Zinc Telluride \\
(CdZnTe) detectors are semiconductor detectors with a function principle similar to SDDs. However, in CdZnTe detectors, the electrons collected in each pixel are read out independently \citep{arnaud_smith_siemiginowska_2011}. Nowadays, CdZnTe detectors become the modern standard instrument for detecting X-rays with energies above several keV, and have been applied to \textit{Swift} and \textit{NuSTAR}.

\section{Pulse TOA estimation}\label{sect:develop_toa}
Given the large distance and so the low X-ray flux of a pulsar, a pulsar cannot be detected to have regularly distributed pulses like a beacon. In this case, a spacecraft can only record a series of events within an exposure, which lasts from $t_{0}$ to $t_{\mathrm{f}}$, on a pulsar \cite{Emadzadeh2010}. It should be noted that an event indicates not only the arrival of an X-ray photon from the pulsar but also possibly the arrival of the background, which could come from CXB, space particles and the nebula surrounding the pulsar.

When there are $M$ events recorded, the events can be denoted as $\{t_{i}\}_{i=1}^{M}$ with $t_{i}$ denoting the arrival time of the $i$th event. As a consequence, the pulse TOA estimation is to estimate the pulse TOA at $t_{0}$ or at $t_{\mathrm{f}}$ out of $\{t_{i}\}_{i=1}^{M}$. In most cases, the pulse TOA estimation is accomplished by estimating the pulse phase.

\subsection{Pulsar signal model}
Given that $t_{i}$ denotes the $i$th event between $\left[t_{0}, t_{\mathrm{f}}\right)$ and that the event indicates the arrival of X-ray photon or the background, each event follows a Poisson distribution \citep{Ross2014}. As a consequence, $\{t_{i}\}_{i=1}^{M}$ is a  nonhomogeneous Poisson process (NHPP) with a time-varying rate $\lambda(t)\geq 0$ \cite{Emadzadeh2010}. $\{t_{i}\}_{i=1}^{M}$ indicates there are $M$ events between  $\left(t_{0}, t_{\mathrm{f}}\right)$, and thus, $M$ is also a Poisson random variable with a probability of \cite{Emadzadeh2010}
\begin{equation}\label{eq.pra}
  P\left(M\right)=\frac{\left(\int_{t_{0}}^{t_{\mathrm{f}}} \lambda(\xi) \mathrm{d}\xi\right)^{M} \exp \left(-\int_{t_{0}}^{t_{\mathrm{f}}} \lambda(\xi) \mathrm{d} \xi\right)}{M !}
\end{equation}

$\lambda(t)$ can be expressed as
\begin{equation}\label{eq.lambda}
  \lambda(t) = \alpha h\left(\phi_{\mathrm{det}}(t)\right)+\beta \quad  \mathrm{counts/s}
\end{equation}
where $h\left(\phi\right)$ is the periodic pulse profile or template, $\phi_{\mathrm{det}}(t)$ is the detected phase, $\alpha$ and $\beta$ are the known pulsed count rate and background count rate, respectively, and counts/s denotes "counts per second".

Assuming the spacecraft is performing a linear uniform motion towards the pulsar or being stationary, $\phi_{\mathrm{det}}(t)$ in Eq.(\ref{eq.lambda}) is expressed as
\begin{equation}\label{eq.phase_model}
  \phi_{\mathrm{det}}(t) = \phi_{0}+f_{0}\left(t-t_{0}\right)
\end{equation}  
where $\phi_{0}$ is the pulse phase at $t_{0}$ and $f_{0}$ is the frequency of the detected pulsar signal. If the spacecraft is stationary, $f_{0}=F_{0}$. When the spacecraft moves towards the pulsar along a line, $f_{0}$ becomes
\begin{equation}\label{eq.f_doppler}
  f_{0} = \left(1+\frac{v}{c}\right)F_{0}
\end{equation} 
\noindent
where $v$ is the velocity of spacecraft along the spacecraft-pulsar line.

Taking the SEXTANT program as an example, Fig.\ref{fig:pulsar_profile} shows the pulse profiles of six pulsars, and Table \ref{tbl:a_b_sextant} provides $\alpha$ and $\beta$ of the pulsars selected by the program. As shown in Table \ref{tbl:a_b_sextant}, the background count rates of all pulsars are at least one order of magnitude higher than the pulsed count rates. Although the Crab pulsar has the highest pulsed count rate among the pulsars, its background count rate is still two orders of magnitude higher than its pulsed count rate. It indicates that all the X-ray pulsars are very faint sources. Moreover, the arrival of photons or background is random, which indicates that the pulse phase estimation belongs to the stochastic signal processing.

\begin{table}[!h]
  \caption{Pulsed count rates ($\alpha$) and background count rates ($\beta$) of pulsars selected by SEXTANT program\citep{ray2017}}\label{tbl:a_b_sextant}
  \begin{tabular*}{\columnwidth}{lrr}
  \toprule
  Pulsar & $\alpha \quad (\mathrm{cnts/s})$ & $\beta\quad (\mathrm{cnts/s})$\\ 
  \midrule
  PSR B1937+21 & 0.029& 0.24\\
  PSR B1821-24 & 0.093& 0.22\\
  PSR J0218+4232 & 0.082& 0.20\\
  PSR J0030+0451 & 0.193& 0.20\\
  PSR J1012+5307 & 0.046& 0.20\\
  PSR J0437-4715 & 0.283& 0.62\\
  PSR J2124-3358 & 0.074& 0.20\\
  PSR J0751+1807 & 0.025& 0.20\\
  PSR J1024-0719 & 0.015& 0.20\\
  PSR J2214+3000 & 0.029& 0.26\\
  Crab pulsar & 660.000 & 13860.20\\
  \bottomrule
  \end{tabular*}
\end{table}

\subsection{Basic pulse phase estimation frames}\label{sect:basic_frame}
The basic pulse phase estimation problem is to estimate $\phi_{0}$ in Eq.(\ref{eq.phase_model}) under the assumption that the spacecraft is performing a uniform linear motion toward the pulsar or stationary and that the evolution of spin frequency of pulsar is ignored.

This section introduces the current methods to solve the above problem. These methods can be divided into two classes: (A) estimating $\phi_{0}$ using epoch folding, and (B) estimating $\phi_{0}$ with the direct use of $\{t_{i}\}_{i=1}^{M}$.

\subsubsection{Pulse phase estimation using epoch folding}

(1) Estimation with known $f_{0}$\label{sect:know_f}

When $f_{0}$ in Eq.(\ref{eq.phase_model}) is already known, the period of pulsar signal equals to $P=1/f_{0}$ and an empirical profile can be recovered from $\{t_{i}\}_{i=1}^{M}$. This recovery method is called epoch folding.

\begin{figure}[h]
  \centering
  \includegraphics[width=\columnwidth]{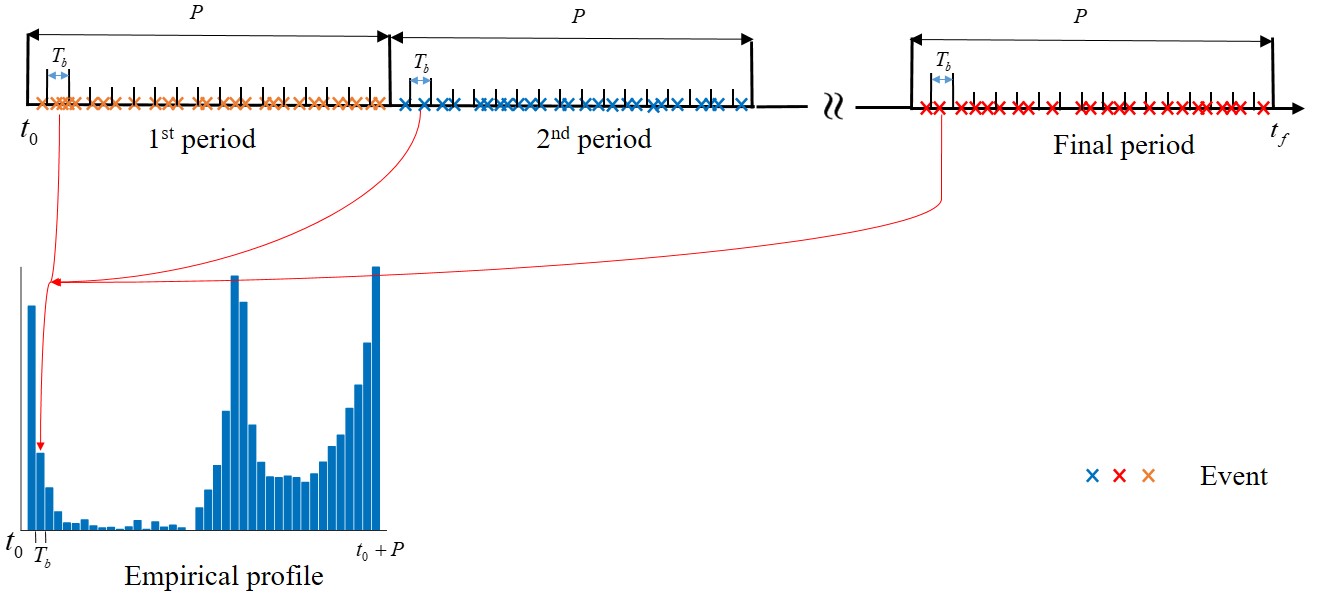}
  \caption{Epoch folding procedure of a pulsar}\label{fig:epoch_folding}
\end{figure}

Assuming the exposure $\left[t_{0}, t_{f}\right)$ contains $N$ periods, the epoch folding consists of three steps: (A) dividing the $i$th ($i=1, 2, \cdots, N$) period into $N_{b}$ bins, each lasting for $T_{b}=P/N_{b}$; (B) the events falling in the $j$th bin of the latter periods are folded back into the $j$th bin of the first period (see Fig.\ref{fig:epoch_folding}); (C) the event counts in each bin are normalized. Finally, an empirical profile is recovered. The statistical properties of the empirical profile have been investigated in Ref.\citenum{Emadzadeh2010}. It should be noted that the epoch folding is applied not only to XNAV but also to X-ray pulsar astronomy \citep{Kirsch2006,Martin-Carrillo2012}. This paper focuses on the application of epoch folding on XNAV.

The phase offset between the recovered empirical profile and the template is called $\phi_{0}$. Thus, the estimation of $\phi_{0}$ can be recast as the classical problem of estimating phase delay between two waveforms. The classical methods to estimate $\phi_{0}$ are cross-correlation \citep{Emadzadeh2011} and Nonlinear Least Square (NLS) \citep{Emadzadeh2010}. The Cramer-Rao Low Bounds (CRLBs) for the $\phi_{0}$s estimated by cross-correlation and NLS are derived in Refs.\citenum{Emadzadeh2010, Emadzadeh2011}.

Motivated by Refs.\citenum{Emadzadeh2010, Emadzadeh2011}, many works started to pursue the fast computation, which is typically accomplished by Fast Fourier Transformation (FFT). For an empirical profile with $N_{b}$ bins, the computational complexity of cross-relation \citep{lin_xu_2019} with the aid of FFT is about $\mathcal{O} \left(N_{b}\log_{N_{b}}\right) $, much less than NLS (about $\mathcal{O} \left(N_{b}^{2}\right)$). FFT is also utilized by Refs.\citenum{XUE2016746} and \citenum{RINAURO2013326}, the bispectrum with the aid of FFT is applied in Ref.\citenum{WU2020163790}, and the Discrete Fourier Transformation (DFT) is employed by Ref.\citenum{lin_xu_2015}.

(2) Estimation with unknown $f_{0}$

When $f_{0}$ in Eq.(\ref{eq.phase_model}) is unknown, we should first estimate $f_{0}$ and then estimate $\phi_{0}$ with the methods discussed above.

The estimation of $f_{0}$ is commonly solved by finding the period of a pulsar. Table \ref{tbl:perid_finding}\citep{Lomb1976, Scargle1982, VanderPlas2018,Zechmeister2009,Huijse2012,Schwarzenberg1989,Schwarzenberg1996,Stellingwerf1978,Palmer2009,Dworetsky1983,Graham2013a,LIU201990,Liu8945227} lists the current period finding algorithms, in which the Lomb–Scargle, Generalized Lomb–Scargle and Correntropy kernel periodogram find the period in the frequency domain, and the other algorithms belong to time-domain methods. Although those methods are derived from different principles, they are dependent on the quality of the light curve of the pulsar. When the exposure on the pulsar is short and the pulsar has a faint magnitude, the quality of the final light curve is low and all those methods become inefficient \citep{Graham2013}.

\begin{table}[htbp]
  \caption{Current period finding algorithms}\label{tbl:perid_finding}
  \begin{tabular*}{\columnwidth}{ll}
  \toprule
  Categroy &Algorithm \\ 
  \midrule
  \multirow{9}{*}{Astronomy} &
  Lomb–Scargle\citep{Lomb1976, Scargle1982, VanderPlas2018}  \\
  &Generalized Lomb–Scargle\citep{Zechmeister2009}  \\
  &Correntropy kernel periodogram\citep{Huijse2012}  \\
  &Binned analysis of variance\citep{Schwarzenberg1989}  \\
  &Multiharmonic analysis of variance\citep{Schwarzenberg1996} \\
  &Phase dispersion minimization\citep{Stellingwerf1978} \\
  &FastChi\citep{Palmer2009} \\
  &String length\citep{Dworetsky1983}\\
  &Conditional entropy\citep{Graham2013a}\\
  \multirow{2}{*}{XNAV} &  Compressive sensing-based method\citep{LIU201990} \\
  &Fast butterfly epoch folding-based method\citep{Liu8945227}  \\
  \bottomrule
  \end{tabular*}
  \end{table}

\subsubsection{Pulse phase estimation with direct use of events}
When $f_{0}$ in Eq.(\ref{eq.phase_model}) is known, according to Eqs.(\ref{eq.pra})-(\ref{eq.phase_model}), the $M$-dimensional joint probability density of $\{t_{i}\}_{i=1}^{M}$ is \citep{Emadzadeh2010}
\begin{equation} \label{eq.MLE}
  p\left(\left\{t_{i}\right\}_{i=1}^{M}, \phi_{0}\right)=\mathrm{exp}\left(-\int_{t_{0}}^{t_{f}} \lambda(\tau;\phi_{0}) \mathrm{d} \tau\right) \prod_{i=1}^{M} \lambda\left(t_{i};\phi_{0}\right)
\end{equation}

Eq.(\ref{eq.MLE}) can be recognized as the likelihood function, and the Log-Likelihood Function (LLF) is
\begin{equation}\label{eq.LLF}
  \operatorname{LLF}\left(\phi_{0}\right)=\sum_{i=1}^{M} \ln \left(\lambda\left(t_{i} ; \phi_{0}\right)\right)-\int_{t_{0}}^{t_{f}} \lambda(\tau;\phi_{0}) \mathrm{d} \tau
\end{equation}

The Maximum Likelihood Estimator (MLE) is provided by maximizing Eq.(\ref{eq.LLF}) with respect to the unknown $\phi_{0}$. The second term on the right side of Eq.(\ref{eq.LLF}) is not sensitive to $\phi_{0}$. As a consequence, the estimate of $\phi_{0}$, $\hat{\phi}_{0}$, can be expressed as
\begin{equation}
  \hat{\phi}_{0}=\underset{\phi_{0} \in(0,1)}{\arg \max } \sum_{i=1}^{M} \ln \left(\lambda\left(t_{i} ; \phi_{0}\right)\right)
\end{equation}
where $\arg \max \left(\bullet\right)$ denotes the argument of the maximum of the function within $\left(\bullet\right)$.

If $f_{0}$ in Eq.(\ref{eq.phase_model}) is unknown, the right side of Eq.(\ref{eq.LLF}) can be viewed as a function of $\phi_{0}$ and $f_{0}$. Then, $\phi_{0}$ and $f_{0}$ can be estimated simultaneously by maximizing Eq.(\ref{eq.LLF}). In this case, we have
\begin{equation}\label{eq.phi_f}
  \hat{\phi}_{0}, \hat{f}_{0}=\arg \max \sum_{i=1}^{M} \ln \left(\lambda\left(t_{i} ; \phi_{0}, f_{0}\right)\right)
\end{equation}

As shown in Eq.(\ref{eq.phi_f}), MLE provides a more flexible frame to estimate $\phi_{0}$ and $f_{0}$ than the epoch folding method. In other words, if the evolution of spin frequency  is taken into consideration, MLE can also work well, and can provide a result closer to the CRLB than NLS \citep{Emadzadeh2010}.

Eq.(\ref{eq.phi_f}) is usually solved by a two-dimensional grid search. When the grid for search is $N_{\phi}\times N_{f}$, the computational complexity of the MLE is about $\mathcal{O} \left(N_{\phi}\times N_{f}\times M\right)$, which is much higher than the computational complexity of NLS even when $N_{b}=N_{\phi}=N_{f}$. In order to accelerate the computation of Eq.(\ref{eq.phi_f}), the intelligent optimization method is employed\citep{Yusong2021}. However, the intelligent optimization method that utilizes the technique of random numbers cannot have a stable estimation result \citep{Yusong2021}.

\subsection{On-orbit pulse phase estimation}
\label{sect:ppe}
The methods introduced in Section \ref{sect:basic_frame} are based on the assumptions that the spacecraft is stationary or performs a linear uniform motion towards the pulsar. However, in practice, spacecraft orbit around a central celestial body, such as the Earth for the Earth-orbiting satellites or the Sun for deep space spacecraft \citep{Battin1999}. In this case, $f_{0}$ in Eq.(\ref{eq.phase_model}) is time-varying and unknown, and then, Eq.(\ref{eq.phase_model}) can be rewritten as \citep{Golshan2007}
\begin{equation}
  \phi_{\mathrm{det}}(t)=\theta_{0}+\int_{t_{0}}^{t} f\left(\tau\right) \mathrm{d} \tau \equiv \theta_{0}+\theta(t)
\end{equation}
\noindent
where
\begin{equation}
  \theta(t)=f_{\mathrm{s}}\left(t-t_{0}\right)+\underbrace{\int_{t_{0}}^{t} f_{\mathrm{d}}\left(\tau\right) \mathrm{d} \tau}_{\equiv \theta_{\mathrm{d}}(t)}
\end{equation}

where $f_{\mathrm{s}}$ is the spin frequency of the pulsar and $\theta_{\mathrm{d}}(t)$ is the phase caused by the Doppler frequency.

In this case, there are currently two frames for estimating the pulse phase: (A) pulse phase tracking and (B) estimation with linearized pulse phase model.

\subsubsection{Pulse phase tracking}
The pulse phase tracking algorithm, which consists of a MLE followed by a {Digital Phase-Locked Loop (DPLL), was first proposed in Ref.\citenum{Golshan2007}. The main idea, illustrated in Fig.\ref{fig:phase_tracking}, is to partition the whole exposure into $N$ blocks and the detected signal frequency is approximately constant over each block. Events obtained over the $i$th block are processed by the two-dimensional MLE shown in Eq.(\ref{eq.phi_f}). In order to reduce the estimation errors or noise in the MLE output sequence, the MLE output sequence is smoothed by the DPLL \citep{Stephens1995}.

\begin{figure}[!htp]
  \centering
  \includegraphics[width=\columnwidth]{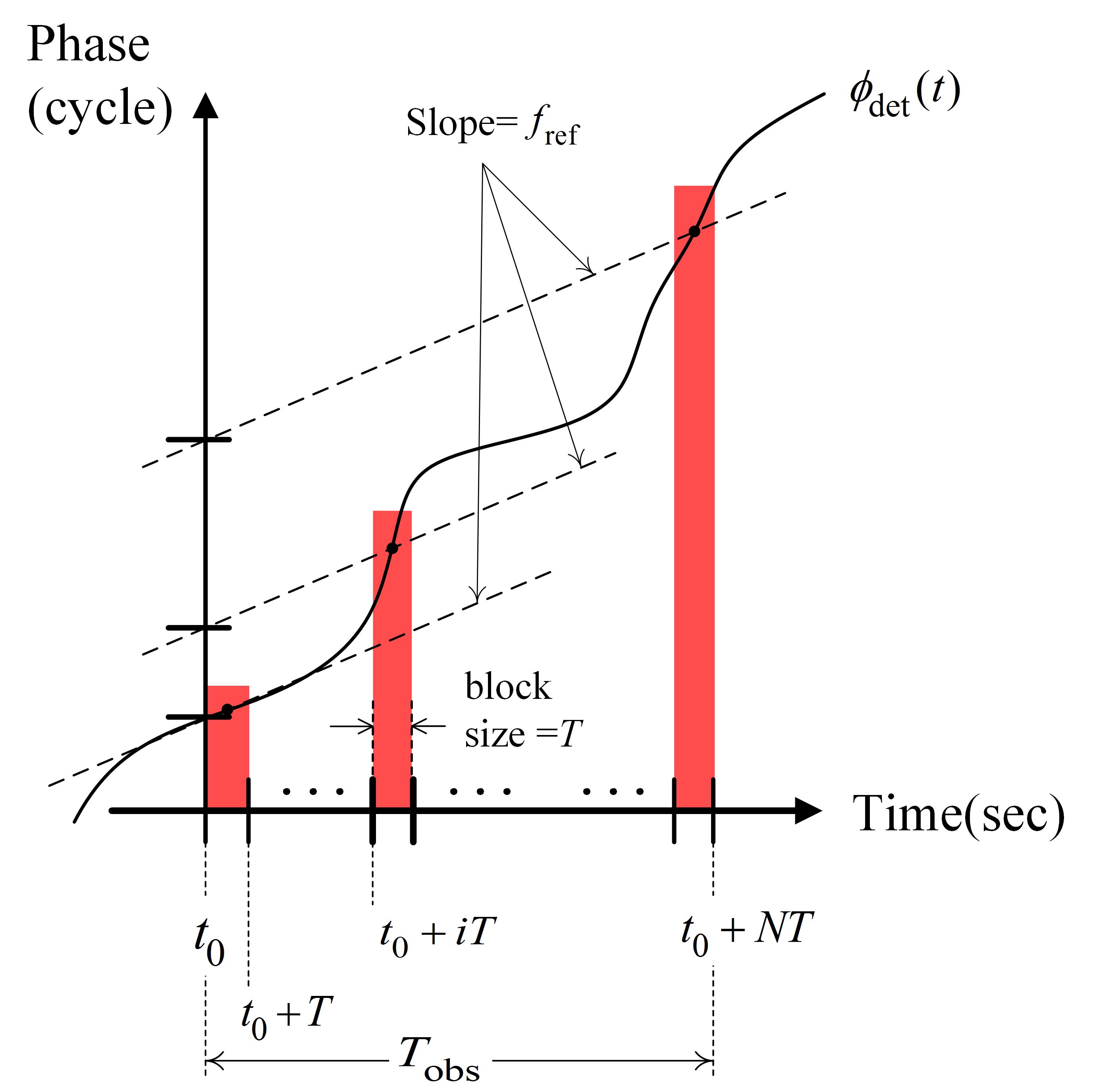}
  \caption{Diagram illustrating the pulse phase tracking concept}\label{fig:phase_tracking}
\end{figure}

Ref.\citenum{Anderson2015} verified the pulse phase tracking algorithm in three cases where the simulated trajectories were all one-dimensional, and relaxed the constant signal frequency assumption using a MLE with a second-order Taylor polynomial phase model and feedback of frequency and its first derivative using a third-order digital phase-locked loop\citep{Anderson2022}. It is found that the phase tracking has a great promise for deep space navigation, but only more limited potential in scenarios where the orbital dynamics is high and the faint MSPs are observed for navigation.

\subsubsection{Pulse estimation with linearized pulse phase model}
As illustrated in Ref.\citenum{Anderson2022}, the phase tracking algorithm limits its potential in the environment with high orbital dynamics such as the Earth space. However, the current flight experiments on XNAV have to be performed on an Earth-orbiting spacecraft. In this case, the pulse estimation with linearized pulse phase model was proposed \citep{Winternitz2016, Luke2018, wang_zheng_2016, Wang2016}.

It follows from Eq.(\ref{BC}) that Eq.(\ref{eq.timing_model}) can be recognized as a function of $\boldsymbol{r}_{\mathrm{SSB}}(t)$. In practice, we can have a prior information on $\boldsymbol{r}_{\mathrm{SSB}}(t)$, denoted as $\tilde{\boldsymbol{r}}_{\mathrm{SSB}}(t)$. Then, Eq.(\ref{eq.timing_model}) can be linearized around $\tilde{\boldsymbol{r}}_{\mathrm{SSB}}(t)$, yielding
\begin{eqnarray}
  \left\{
  \begin{aligned}
    &\phi\left(\boldsymbol{r}_{\mathrm{SSB}}(t)\right) = \phi\left(\tilde{\boldsymbol{r}}_{\mathrm{SSB}}(t)\right)+\left.\frac{\partial \phi}{\partial \boldsymbol{r}}\right|_{ \boldsymbol{r}=\tilde{\boldsymbol{r}}_{\mathrm{SSB}}(t)}\delta \boldsymbol{r}_{\mathrm{SSB}}(t)\\
    &\delta \boldsymbol{r}_{\mathrm{SSB}}(t)=\boldsymbol{r}_{\mathrm{SSB}}(t)-\tilde{\boldsymbol{r}}_{\mathrm{SSB}}(t)
  \end{aligned}
    \right.
\end{eqnarray}
$\tilde{\boldsymbol{r}}_{\mathrm{SSB}}(t)$ can be predicted from $\tilde{\boldsymbol{r}}_{\mathrm{SSB}}(t_{0})$, obtained by other navigation methods or by propagating from the last epoch, by propagating the orbit dynamics model of spacecraft \citep{Battin1999}. As a consequence, $\phi\left(\boldsymbol{r}_{\mathrm{SSB}}(t)\right)$ can be expressed as \citep{Winternitz2016, Luke2018}
\begin{equation}
  \phi\left(\boldsymbol{r}_{\mathrm{SSB}}(t)\right)=\phi\left(\tilde{\boldsymbol{r}}_{\mathrm{SSB}}(t)\right)+\phi_{0}+f_{\mathrm{a}}\left(t-t_{0}\right)
\end{equation}
\noindent
where $\phi_{0}$ and $f_{\mathrm{a}}$ are unknown.

Given that the events set $\{t_{i}\}_{i=1}^{M}$ follows a Poisson process,\\ $\{\phi\left(\boldsymbol{r}_{\mathrm{SSB}}(t_{i})\right)\}_{i=1}^{M}$ also follows a Poisson process  with constant $\phi_{0}$ and $f_{\mathrm{a}}$. Thus, $\phi_{0}$ and $f_{\mathrm{a}}$ can be estimated by solving a two-dimensional optimization problem:
\begin{equation}\label{eq.phi_f2}
  \hat{\phi}_{0}, \hat{f}_{\mathrm{a}}=\arg \max \sum_{i=1}^{M} \ln \left(\lambda\left(\phi_{0}, f_{\mathrm{a}}, \tilde{\boldsymbol{r}}_{\mathrm{SSB}}(t_{i})\right)\right)
\end{equation}

The SEXTANT team employs the two-dimensional grid search to solve the above problem \citep{Winternitz2016}. In order to reduce the computational complexity, the parallel computation is employed to accelerate the grid search \citep{wang_zheng_2016}, and decouples the estimation of $\phi_{0}$ and $f_{\mathrm{a}}$ for deep space exploration missions \citep{Wang2016}. $f_{\mathrm{a}}$ is first searched out and then $\phi_{0}$ is estimated by epoch folding. Recently, an on-orbit pulsar timing analysis is proposed to fast estimate $\phi_{0}$, and has been verified by the data from XNAV flight experiments\citep{Wang2022}.

\section{Methods to improve the navigation performance of XNAV}\label{sect:p_app_xnav}
The principle of XNAV was briefly summarized in Section \ref{sect:p_XNAV}. We first need to find an optimal pulsar combination from a given navigation pulsar database. When the optimal pulsar combination is achieved, both the geometry for a specific navigation mission and the corresponding navigation performance of XNAV in this case can be improved. Moreover, there are many systematic biases within XNAV, which limit the navigation performance of XNAV. Furthermore, the conventional XNAV works with the assumption that a single spacecraft should load at least three X-ray detection systems with large areas to receive the pulsar signal from three directions. This assumption is too ideal for the practical applications. Therefore, in order to improve the navigation performance of XNAV, this section will introduce three types of methods.

\subsection{Optimal pulsar combination}\label{sect:opc}
This work is usually accomplished by investigating the observability of the navigation system. Most time, the observability is evaluated by the determinant of Fisher Information Matrix (FIM). When there are three pulsars being observed simultaneously, the FIM is derived as \citep{WANG201768}
\begin{equation} \label{eq.FIM}
  \boldsymbol{F}=\sum_{i=1}^{3} \sigma_{i}^{-2} \boldsymbol{n}_{i} \boldsymbol{n}_{i}^{\mathrm{T}}
\end{equation}
where $\boldsymbol{n}_{i}$ is the direction vector of the $i$th pulsar and $\sigma_{i}$ is the accuracy of the $i$th pulse TOA.

Based on Eq.(\ref{eq.FIM}), Ref.\citenum{YU2015136} analytically investigates the optimal pulsar combination, and find that the determinant of Eq.(\ref{eq.FIM}) will reach its maximum if $\boldsymbol{n}_{1}$, $\boldsymbol{n}_{2}$ and $\boldsymbol{n}_{3}$ are orthogonal to each
other. This conclusion can be extended to a general case where a spacecraft observes $N$ pulsars at the same time. In this case, an optimal pulsar combination should be composed of pulsars with direction vectors orthogonal to each other.

\subsection{Systematic biases analysis and compensation}
As shown in Eq.(\ref{BC}), the errors within the direction vector of pulsar, positions of planets and the atomic clock recording arrival time of events would affect the navigation performance of XNAV.

For the autonomous navigation in Earth orbits, the impacts of errors within the direction vector of the pulsar, errors within the ephemerides of planets, and clock errors have been analyzed\citep{LIU20101409, XU20183187, WANG2013, WANG201427}. It is found that those errors vary according to the orbit period of the Earth and thus can be approximated as constants for a navigation process with duration much less than one orbit period of the Earth. This conclusion paves the way for compensating the impacts of those errors. When the XNAV is applied to multi-spacecraft, it has been found that the systematic biases can be eliminated when all the spacecraft observe a common pulsar simultaneously \citep{KAI2009427}.

The compensation methods generally consist of three categories, including the augmented-state method \citep{LIU20101409}, the time-differenced method \citep{WANG201427, Wang2019} and the $H_{\infty}/H_{2}$ filter \citep{Xiong2010}. The augmented-state method augments the systematic biases into the state vector, and estimates the systematic biases along with the position and velocity of spacecraft. However, when the number of systematic biases is more than 6, the number of position and velocity, Kalman-type filters employed for estimation might be unstable. Thus, the time-differenced method was proposed. The method employs the difference between pulse TOAs at adjacent time as the measurement for estimation, avoiding the high-dimensional matrices that were involved in the augmented-state method. The $H_{\infty}/H_{2}$ filter, originating from the $H_{\infty}$ control, derives an upper bound for the systematic biases, and suppresses the impacts of systematic biases by the bounds. However, how to design a proper upper bound is still an open question.

\subsection{Integrated navigation}
Up to now, the information to be combined with XNAV roughly includes the Earth image information obtained from ultraviolet sensor \citep{Qiao5290372}, the Doppler velocity measurements relative to the Sun \citep{Wang2015} or the Mars \citep{CUI20161889}, the optical measurements on the Mars \citep{liu_wei_jin_2017, GU2019512} or the Earth \citep{wang_zheng_an_sun_li_2013, KAI2016473}, and the output of inertial navigation system \citep{XU201828, Wang2016a}. For the integrated navigation system fusing the star angle information and XNAV, the impacts of direction vector of pulsar\citep{Ning7973088}, and the ephemerides errors of Jupiter and other planets\citep{NING201736, Gui8520886, Ning2018} have been investigated. The integrated navigation is applied not only to a single spacecraft but also to multi-spacecraft \citep{xin_wang_zheng_meng_zhang_2018, Liu7081637}.

The fusion of information from other sources and XNAV is typically achieved by the federated Kalman filter. The federated Kalman filter consisting of parallel filters for each information fuses the outputs of the parallel filters to obtain the final estimation result. Given that the parallel filters work independently, the final estimation result is suboptimal  \citep{wang_zheng_an_sun_li_2013}. A nonlinear kinematic and static filter is derived \citep{wang_zheng_an_sun_li_2013, Wang2016a} to overcome the suboptimal fusion of federated Kalman filter. A step Kalman filter structure is designed\citep{SHANG201333} to optimally fuse the parallel filters that have greatly different filtering precision. However, the above information fusions work by fusing the pulse phase or TOA and the measurements from the other sources. The achievements of such information fusions are based on the assumption that the pulse phase is obtained. However, as illustrated in Section \ref{sect:ppe}, the on-orbit pulse TOA estimation problem has not been well solved. It is a good idea to employ the measurements from the other sources to estimate the pulse phase\citep{KAI2016473,wang_zheng_zhang_2017}.

\section{Conclusions and future work}
In this paper, we review X-ray pulsar navigation (XNAV) systems and the algorithms, and briefly introduce the past, present and future missions with XNAV experiments. This paper focuses on the advances of the key techniques supporting XNAV, including the navigation pulsar database, the X-ray detection system, and the pulse TOA estimation, as well as the methods to improve the navigation performance of XNAV.

It can be learned from the review that although several flight experiments on XNAV have been successfully complemented, the wide applications of XNAV still have a long way to go. Further development of XNAV may be pursued from the following aspects:

\begin{enumerate}[(1)]
  \item Designing X-ray detection systems dedicated for navigation. The current X-ray detection systems in space, serve for X-ray astronomy missions, which pursue high-quality images or signals with high signal-to-noise ratios. In this case, the volume and weight of an X-ray detection system dominate the payload. However, for a spacecraft with XNAV, the X-ray detection system is just a subsystem of the whole spacecraft. As a consequence, an X-ray detection system dedicated for navigation should be of a small size but have a high detection performance.
  \item Pursuing computationally efficient on-orbit pulse TOA estimation method. Currently, only the pulse TOA estimation for millisecond pulsars have been successfully complemented in orbit. The high flux of Crab pulsar makes the XNAV with small X-ray detection system become applicable, although the stability of Crab pulsar is far less than millisecond pulsars. However, how to quickly estimate the pulse TOA of the Crab pulsar in orbit is an open question. It is because the computational burden in this case is much higher than millisecond pulsars.
  \item Verifying the current navigation algorithms with real X-ray pulsar data. Section \ref{sect:p_app_xnav} reviews various navigation algorithms that have been proposed to improve the navigation performance of XNAV. However, most of those algorithms work with simulation data under the assumption that the high-precision pulse ToAs can be obtained. In the algorithms, X-ray detection systems are assumed to have an area of about 1 $\mathrm{m}^{2}$. However, as shown in Tables \ref{tbl:satellite_collimator} and \ref{tbl:wolter}, such a big area is not easy to achieve in space, even for dedicated X-ray astronomical missions. Furthermore, the flux of a pulsar's pulsed signal is much less than the background caused by the particles and X-rays from other sources. Thus, the pulse TOA estimation result in the real environment is less accurate than the values assumed in the navigation algorithms. As a consequence, the performance of those navigation algorithms in real cases is quite probably not as good as that in simulations.
\end{enumerate}

\section*{Acknowledgements}

This work is funded by the National Natural Science Foundation of China (No. 61703413) and the Science and Technology Innovation Program of Hunan Province (No. 2021RC3078).

\nocite{*}









\end{document}